\documentclass[%
 aip,
 jmp,%
 amsmath,amssymb,floatfix,
 reprint,%
]{revtex4-1}

\usepackage[utf8]{inputenc}
\usepackage[T1]{fontenc} 

\usepackage[english]{babel}

\usepackage{graphicx}
\usepackage{dcolumn}
\usepackage{bm}

\usepackage{amsmath}
\usepackage{amsfonts}
\usepackage{amssymb}
\usepackage{braket}
\usepackage{xcolor}

\usepackage{lmodern}

\usepackage[colorlinks=true,linkcolor=blue]{hyperref}

\newcommand{\vect}[1]{\boldsymbol{#1}}

\newcommand{\boldNabla}{\boldsymbol{\nabla}}

\newcommand{\kS}{\kappa_\mathrm{S}}
\newcommand{\kA}{\kappa_\mathrm{A}}
\newcommand{\kB}{\kappa_\mathrm{B}}

\newcommand{\kBolt}{k_\mathrm{B}}

\newcommand{\alphaB}{\alpha_\mathrm{B}}

\newcommand{\betaB}{\beta_\mathrm{B}}

\newcommand{\TS}{T_\mathrm{S}}
\newcommand{\TB}{T_\mathrm{B}}

\newcommand{\G}{\mathcal{G}}
\newcommand{\R}{\vect{r}}
\newcommand{\Intd}{\mathrm{d }}

\newcommand{\F}{\vect{F}}

\newcommand{\Y}{\vect{y}}
\newcommand{\X}{\vect{x}}

\newcommand{\pt}{\tilde{p}}
\newcommand{\vt}{\tilde{v}}
\newcommand{\ut}{\tilde{u}}

\newcommand{\Gt}{\tilde{\G}}
\newcommand{\Ft}{\tilde{F}}

\newcommand{\tauS}{\tau_{\mathrm{S}}}
\newcommand{\tauB}{\tau_{\mathrm{B}}}

\newcommand{\E}{\operatorname{E}}
\newcommand{\bigO}{\mathcal{O}}

\newcommand{\muS}{\mu^{\mathrm{S}}}
\newcommand{\muP}{\mu^{\mathrm{P}}}

\newcommand{\phiPerpShearSelf}{\phi_{zz, \mathrm{S}}^{\mathrm{S}}}
\newcommand{\phiPerpBendingSelf}{\phi_{zz, \mathrm{B}}^{\mathrm{S}}}

\newcommand{\Sm}{S_{\mathrm{m}}}
\newcommand{\Sp}{S_{\mathrm{p}}}

\newcommand{\Faxen}{Fax\'{e}n}

\newcommand{\ETH}{\epsilon_{\mathrm{th}}}

\newcommand{\zetaXXS}{\zeta_{xx}^{\mathrm{S}}}
\newcommand{\zetaYYS}{\zeta_{yy}^{\mathrm{S}}}
\newcommand{\zetaZZS}{\zeta_{zz}^{\mathrm{S}}}

\newcommand{\zetaXXP}{\zeta_{xx}^{\mathrm{P}}}
\newcommand{\zetaYYP}{\zeta_{yy}^{\mathrm{P}}}
\newcommand{\zetaZZP}{\zeta_{zz}^{\mathrm{P}}}

\newcommand{\zetaXZP}{\zeta_{xz}^{\mathrm{P}}}

\begin{document}

\preprint{AIP/123-QED}

\title[Hydrodynamic interaction near elastic interfaces]
{Hydrodynamic interaction between particles near elastic interfaces}

\author{Abdallah Daddi-Moussa-Ider}
\email{abdallah.daddi-moussa-ider@uni-bayreuth.de}

\author{Stephan Gekle}
\affiliation{%
Biofluid Simulation and Modeling, Fachbereich Physik, Universit\"at Bayreuth, \\ Universit\"{a}tsstra{\ss}e 30, Bayreuth 95440, Germany
}%

\date{\today}

\begin{abstract} 

We present an analytical calculation of the hydrodynamic interaction between two spherical particles near an elastic interface such as a cell membrane. The theory predicts the frequency dependent self- and pair-mobilities accounting for the finite particle size up to the 5th order in the ratio between particle diameter and wall distance as well as between diameter and interparticle distance. We find that particle motion towards a membrane with pure bending resistance always leads to mutual repulsion similar as in the well-known case of a hard-wall. In the vicinity of a membrane with shearing resistance, however, we observe an attractive interaction in a certain parameter range which is in contrast to the behavior near a hard wall. This attraction might facilitate surface chemical reactions. Furthermore, we show that there exists a frequency range in which the pair-mobility for perpendicular motion exceeds its bulk value, leading to short-lived superdiffusive behavior. Using the analytical particle mobilities we compute collective and relative diffusion coefficients. The appropriateness of the approximations in our analytical results is demonstrated by corresponding boundary integral simulations which are in excellent agreement with the theoretical predictions.

\end{abstract}


\maketitle

\section{Introduction}

The hydrodynamic interaction between particles moving through a liquid is essential to determine the behavior of colloidal suspensions \cite{Morrone_2012}, polymer solutions \cite{usta05, wojciechowski10}, chemical reaction kinetics \cite{vonHansen_2010, Diugosz_2012}, bilayer assembly \cite{Ando_2013} or cellular flows \cite{Popel_2005, Misbah_2013}.
As an example, hydrodynamic interactions result in a notable alteration of the collective motion behavior of catalytically powered self-propelled particles \cite{molina13} or bacterial suspensions \cite{wensink12, dunkel13, lopez14, Zottl_2014, elgeti15}.
Many of the occurring phenomena can be explained on the basis of two-particle interactions \cite{Guazzelli_2012} which in bulk are well understood. Some of the most intriguing observations, however, are made when particles interact hydrodynamically in the close vicinity of interfaces -- a prominent example being the attraction of like-charged colloid particles during their motion away from a hard wall \cite{larsen97, squires00, behrens01}.


In the low Reynolds number regime hydrodynamic interactions between two particles are fully described by the mobility tensor which provides a linear relation between the force applied on one particle and the resulting velocity of either the same or the neighboring particle. 
In an unbounded flow, algebraic expressions for the hydrodynamic interactions between two \cite{felderhof77, kim84, yoon87, cichocki88a, happel12, Guazzelli_2012} and several \cite{deutch71, batchelor76, ermak78, ladd88, ekiel15, zia15} spherical particles are well established. 
Experimentally, the predicted hydrodynamic coupling has been confirmed using optical tweezers \cite{crocker97, meiners99, bartlett01, henderson02} and atomic force microscopy \cite{radiom15}.


The presence of an interface is known to drastically alter the hydrodynamic mobility. 
For a single particle, this wall-induced drag effect has been studied extensively over recent decades theoretically and numerically near a rigid \cite{lauga05, lorentz07, mackay61, gotoh82, cichocki98, franosch09,felderhof12, padding14, decorato15, huang15}, a fluid-fluid \cite{lee79, berdan81, bickel06, bickel07, blawz10theory, blawz10} or an elastic interface \cite{felderhof06, Shlomovitz_2013, Shlomovitz_2014, salez15, daddi16, daddi16b, saintyves16}. 
While rigid interfaces in general simply lead to a reduction of particle mobility, the memory effect caused by elastic interfaces leads to a frequency dependence of the particle mobility and can cause novel phenomena such as transient subdiffusion \cite{daddi16}. 
On the experimental side, the single particle mobility has been investigated using optical tweezers \cite{faucheux94, dufresne01, schaffer07}, evanescent wave dynamic light scattering \cite{holmqvist07, michailidou09, Wang_2009_diffusion, Kazoe_2011, lisicki12, rogers12, Michailidou_2013, Wang_2014_diffusion, Watarai_2014, lisicki14} or video microscopy \cite{Eral_2010,CervantesMartinez_2011, Dettmer_2014}.
The influence of a nearby elastic cell membrane has recently been investigated using magnetic particle actuation \cite{irmscher12} and optical traps \cite{Shlomovitz_2013, Boatwright_2014, Junger15}.

Hydrodynamic interactions between two particles near a planar rigid wall have been studied theoretically \cite{squires00, swan07} and experimentally using optical tweezers \cite{lele11, trankle16} and digital video microscopy \cite{dufresne00}. 
Narrow channels \cite{Cui_2002, misiunas15}, 2D confinement \cite{Bleibel_2014}
or liquid-liquid interfaces have also been investigated \cite{zhang13, zhang14}.
Near elastic interfaces, however, no work regarding hydrodynamic interactions has so far been reported. Given the complex behavior of a single particle near an elastic interface (caused by the above-mentioned memory effect) such hydrodynamic interactions can be expected to present a very rich phenomenology.


In this paper, we calculate the motion of two spherical particles positioned above an elastic membrane both analytically and numerically.
We find that the shearing and bending related parts in the pair-mobility can in some situations have opposite contributions to the total mobility. Most prominently, we find that two particles approaching an idealized membrane exhibiting only shear resistance will be attracted to each other which is just opposite to the well-known hydrodynamic repulsion  for motion towards a hard wall \cite{squires00}.
Additionally, we show that the pair-mobility at intermediate frequencies may even exceed its bulk value, a feature which is not observed in bulk or near a rigid wall. 
This increase in pair-mobility results in a short-lived superdiffusion in the joint mean-square displacement.

The remainder of the paper is organized as follows.
In Sec.~\ref{sec:theory}, we introduce the theoretical approach to computing the frequency-dependent self- and pair-mobilities from the multipole expansion and \Faxen's theorem, up to the 5th order in the ratio between particle radius and particle-wall or particle-particle distance.
In Sec.~\ref{sec:bim}, we present the boundary integral method which we have used to numerically confirm our theoretical predictions.
In Sec.~\ref{sec:results}, we provide analytical expressions of the particle self- and pair-mobilities in terms of power series, finding excellent agreement with our numerical simulations. 
Expressions of the self- and pair-diffusion coefficients are derived in Sec.~\ref{sec:diffusion}.
Concluding remarks are made in Sec.~\ref{sec:conclusions}.


\section{Theory} \label{sec:theory}

\begin{figure}
  \begin{center}
     \includegraphics[scale = 0.4]{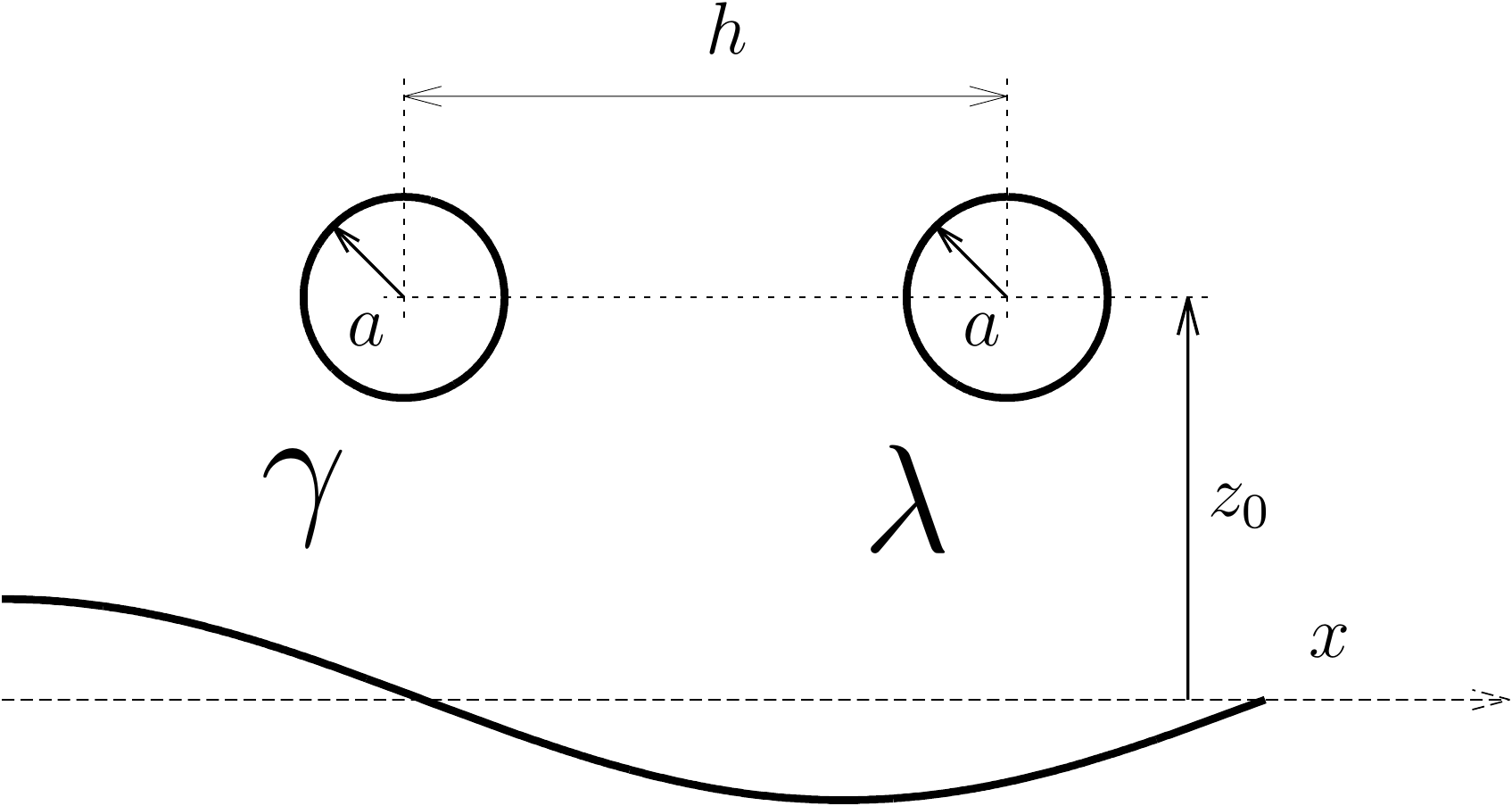}
     \caption{ Illustration of the problem setup. Two small particles labeled $\gamma$ and $\lambda$ of radius $a$ are located a distance $h := x_{\lambda} - x_{\gamma}$ apart and a distance $z_0$ above an elastic membrane. The dimensionless length scales of the problem are $\epsilon:=a/z_0$ and $\sigma:=a/h$.}
     \label{illustration}
  \end{center}
\end{figure}

We consider a pair of particles of radius $a$ suspended in a Newtonian fluid of viscosity $\eta$ above a planar elastic membrane extending in the $xy$ plane.
The two particles are placed at $\R_{\gamma} = (x_{\gamma},0,z_0)$ and $\R_{\lambda} = (x_{\lambda},0,z_0)$, i.e.\ the line connecting the two particles is parallel to the undisplaced membrane.
We denote by $h:= x_{\lambda} - x_{\gamma}$ the center-to-center separation measured from the left ($\gamma$) to the right ($\lambda$) particle (see Fig.~\ref{illustration} for an illustration).

The particle mobility is a tensorial quantity that linearly couples the velocity ${V_{\gamma}}_\alpha$ of particle~$\gamma$ in direction $\alpha$ to an external force in the direction $\beta$ applied on the same (${F_{\gamma}}_\beta$) or the other (${F_{\lambda}}_\beta$) particle.
Transforming to the frequency domain we thus have \cite[ch. 7]{kim05}
\begin{equation}
 {V_{\gamma}}_{\alpha} (\omega) = \mu_{\alpha \beta}^{\gamma \gamma} (\R_{\gamma}, \R_{\gamma},  \omega) {F_{\gamma}}_{\beta} (\omega)
 + \mu_{\alpha \beta}^{\gamma \lambda} (\R_{\gamma}, \R_{\lambda},  \omega) {F_{\lambda}}_{\beta} (\omega)\, , 
 \notag
\end{equation}
where Einstein's convention for summation over repeated indices is assumed.
The particle mobility tensor in the present geometry can be written as an algebraic sum of two distinct contributions
\begin{equation}
 \mu_{\alpha \beta}^{\gamma \lambda} (\R_{\gamma}, \R_{\lambda},  \omega) = {b}_{\alpha \beta}^{\gamma \lambda} (\R_{\gamma}, \R_{\lambda}) 
								  + \Delta \mu_{\alpha \beta}^{\gamma \lambda} (\R_{\gamma}, \R_{\lambda},  \omega) \, ,
\label{eqn:defMobility}								  
\end{equation}
where ${b}_{\alpha \beta}^{\gamma \lambda}$ is the pair-mobility in an unbounded geometry (bulk flow), and $\Delta \mu_{\alpha \beta}^{\gamma \lambda}$ is the frequency-dependent  correction due to the presence of the elastic membrane. An analogous relation holds for $\mu_{\alpha \beta}^{\gamma \gamma}$.

For the determination of the particle mobility, 
we consider a force density $\vect{f}$ acting on the surface $S_{\lambda}$ of the particle~$\lambda$, related to the total force by
\begin{equation}
 {{F}_{\lambda}}_{\beta} (\omega) = \oint_{S_{\lambda}} {f}_{\beta} (\R', \omega) \Intd^2 \R' \, ,
 \notag
\end{equation}
which induces the disturbance flow velocity at point $\R$
\begin{equation}
 v_{\alpha} (\R, \R_{\lambda}, \omega) = \oint_{S_{\lambda}} \G_{\alpha \beta} (\R, \R', \omega) f_{\beta} (\R', \omega) \Intd^2 \R' \, ,
 \label{fluidVelocityIntPointForce}
\end{equation}
where $\G_{\alpha \beta} $ denotes the velocity Green's function (Stokeslet), i.e.\ the flow velocity field resulting from a point-force acting on $\R_{\lambda}$.
The disturbance velocity at any point $\R$ can be split up into two parts,
\begin{equation}
 v_{\alpha} (\R,\R_{\lambda}, \omega) = {v}_{\alpha}^{(0)} (\R, \R_{\lambda}) + \Delta v_{\alpha} (\R, \R_{\lambda}, \omega) \, ,
 \label{fluidVelocitySplitUp}
\end{equation}
where ${v}_{\alpha}^{(0)}$ is the flow field induced by the particle~$\lambda$ in an unbounded geometry, and $\Delta v_{\alpha}$ is the flow satisfying the no-slip boundary condition at the membrane.
In this way, the Green's function can be written as
\begin{equation}
 \G_{\alpha \beta} (\R, \R', \omega) = \G_{\alpha \beta}^{(0)} (\R, \R') + \Delta \G_{\alpha \beta} (\R, \R', \omega) \, ,
 \label{eqn:defDeltaG}
\end{equation}
where $\G_{\alpha \beta}^{(0)}$ is the infinite-space Green's function (Oseen's tensor) given by
\begin{equation}
 \G_{\alpha \beta}^{(0)} (\R,\R') = \frac{1}{8\pi\eta} \left( \frac{\delta_{\alpha \beta}}{s} + \frac{{s}_\alpha {s}_\beta}{s^3} \right) \, ,
 \label{infiniteSpaceGreensFunction}
\end{equation}
with $\vect{s} :=\R-\R'$ and  $s:=|\vect{s}|$.
The term $\Delta \G_{\alpha\beta}$ represents the frequency-dependent correction due to the presence of the membrane.
Far away from the particle~$\lambda$, the vector $\R'$ in Eq.~\eqref{fluidVelocityIntPointForce} can be expanded around the particle center $\R_{\lambda}$ following a multipole expansion approach. 
Up to the second order, and assuming a constant force density, the disturbance velocity can be approximated by \cite{kim06, gauger08, swan07, swan10}
\begin{equation}
  v_\alpha (\R, \R_{\lambda},\omega) \approx \left( 1+\frac{a^2}{6} \boldsymbol{\nabla}_{\R_{\lambda}}^2  \right) \G_{\alpha \beta} (\R, \R_{\lambda}, \omega) {F_{\lambda}}_{\beta} (\omega) \, ,
 \label{flowField_MultipoleExp}
\end{equation}
where $\boldsymbol{\nabla}_{\R_{\lambda}}$ stands for the gradient operator taken with respect to the singularity position $\R_\lambda$.
Note that for a single sphere in bulk, the flow field given by Eq.~\eqref{flowField_MultipoleExp} satisfies exactly the no-slip boundary conditions at the surface of the sphere, i.e.\ in the frame moving with the particle, both the normal and tangential velocities vanish.
Using \Faxen's theorem, the velocity of the second particle~$\gamma$ in this flow reads \cite{kim06, gauger08, swan07, swan10}
\begin{equation}
 {V_{\gamma}}_{\alpha} (\omega) = \mu_0 {{F}_{\gamma}}_{\alpha} (\omega) + \left( 1+\frac{a^2}{6} \boldsymbol{\nabla}_{\R_{\gamma}}^2 \right) v_{\alpha} (\R_{\gamma}, \R_{\lambda}, \omega)  \, ,
 \label{Faxen}
\end{equation}
where $\mu_0 := 1/(6\pi\eta a)$ denotes the usual bulk mobility, given by the Stokes' law.
The disturbance flow $v_{\alpha}$ incorporates both the disturbance from the particle~$\lambda$ and the disturbance caused by the presence of the membrane.
By plugging Eq.~\eqref{flowField_MultipoleExp} into \Faxen's formula given by Eq.~\eqref{Faxen}, the $\alpha\beta$ component of the frequency-dependent pair-mobilities  can be obtained from
\begin{equation}
  \mu_{\alpha\beta}^{\gamma \lambda} (\omega) = \left( 1+\frac{a^2}{6} \boldsymbol{\nabla}_{\R_{\gamma}}^2 \right) 
		       \left( 1+\frac{a^2}{6} \boldsymbol{\nabla}_{\R_{\lambda}}^2  \right) \G_{\alpha\beta} (\R_{\gamma}, \R_{\lambda}, \omega) 
		       \, .
		       \label{particlePairMobility}  
\end{equation}

For the self-mobilities, only the correction in the flow field $\Delta v_{\alpha}$ due to the presence of the membrane in Eq.~\eqref{fluidVelocitySplitUp} is considered in \Faxen's formula (the influence of the second particle on the self-mobility is neglected here for simplicity \cite{swan07, gauger08}). Therefore, the frequency-dependent self-mobilities read
\begin{equation}
 \begin{split}
  \mu_{\alpha\beta}^{\gamma \gamma} (\omega) &= \mu_0
					     + \lim_{\R \to \R_\gamma} \left( 1+\frac{a^2}{6} \boldsymbol{\nabla}_{\R}^2 \right)  \\
		       &\quad\times \left( 1+\frac{a^2}{6} \boldsymbol{\nabla}_{\R_{\gamma}}^2  \right) \Delta \G_{\alpha\beta} (\R, \R_{\gamma}, \omega) \, 
		       \label{particleSelfMobility}
\end{split}
\end{equation}
and analogously for $\mu_{\alpha\beta}^{\lambda\lambda}$.


In order to use the particle pair- and self-mobilities from Eqs.~\eqref{particlePairMobility} and \eqref{particleSelfMobility}, the velocity Green's functions in the presence of the membrane are required. 
These have been calculated in our earlier work \cite{daddi16} and their derivation is only briefly sketched here with more details in Appendix~\ref{appendix:derivationGreenFcts}.


We proceed by solving the steady Stokes equations with an arbitrary time-dependent point-force $\vect{F}$ acting at $\R_0 = (0, 0, z_0)$,
\begin{align}
 \eta \boldNabla^2 \vect{v} - \boldNabla p  + \F \delta (\R-\R_0) &= 0 \, , \label{eq:momentum}\\
 \boldNabla \cdot \vect{v} &= 0 \, , \label{eq:incompressibility}
\end{align}
where $p$ is the pressure field.
The determination of the Green's functions at $\R_{\lambda}$ is straightforward thanks to the system translational symmetry along the $xy$ plane.
After solving the above equations and appropriately applying the boundary conditions at the membrane, we find that the Green's functions are conveniently expressed by \begin{subequations}
\begin{align}
{\G}_{zz} (\R, \R_\lambda, \omega) &= \frac{1}{2\pi}
  \int_{0}^{\infty}  \Gt_{zz} (q,z,z_0,\omega) J_0 (\rho q) q \Intd q \, , \label{Gzz} \\
\G_{xx} (\R, \R_\lambda, \omega) &= \frac{1}{4\pi}  \int_0^{\infty} \bigg( \Gt_{+} (q,z,z_0,\omega) J_0 (\rho q) \notag \\
   &+   \Gt_{-} (q,z,z_0,\omega) J_2 (\rho q) \cos 2\theta \bigg) q \Intd q \, ,  \label{Gxx} \\
\G_{yy} (\R, \R_\lambda,\omega) &= \frac{1}{4\pi}  \int_0^{\infty} \bigg( \Gt_{+} (q,z,z_0,\omega) J_0 (\rho q) \notag \\
   &-   \Gt_{-} (q,z,z_0,\omega) J_2 (\rho q) \cos 2\theta \bigg) q \Intd q \, ,  \label{Gyy} \\
\G_{xz} (\R,\R_\lambda, \omega) &= \frac{i \cos \theta}{2\pi} \int_{0}^{\infty} \Gt_{lz} (q,z,z_0,\omega) J_1 (\rho q) \Intd q \, , \label{Gxz}
\end{align}
\end{subequations}
where $\rho:= \sqrt{(x-x_\lambda)^2 + y^2}$, $\theta := \arctan (y/(x-x_\lambda))$ with $\R=(x,y,z)$.
Here $J_n$ denotes the Bessel function of the first kind of order $n$.
The functions $\Gt_{\pm}$, $\Gt_{lz}$ and $\Gt_{zz}$  are provided in Appendix~\ref{appendix:derivationGreenFcts}.
It is worth to mention here that the unsteady term in the Stokes equations leads to negligible contribution in the correction to the Green's functions  \cite{daddi16}, and it is therefore not considered in the present work.

The membrane elasticity is described by the well-established Skalak model \cite{skalak73}, commonly used to describe deformation properties of red blood cell (RBC) membranes \cite{eggleton98, krueger11, krueger12}.
The elastic model has as parameters the shearing modulus $\kS$ and the area-expansion modulus $\kA$.
The two moduli are related via the dimensionless number $C := \kA/\kS$.
Moreover, the membrane resists towards bending according to Helfrich's model \cite{helfrich73}, with the corresponding bending rigidity $\kB$.


\section{Boundary integral methods}\label{sec:bim}

In this section, we introduce the numerical method used to compute the particle self- and pair- mobilities.
The numerical results will subsequently be compared with the analytical predictions presented in Sec.~\ref{sec:theory}.

For solving the fluid motion equations in the inertia-free Stokes regime, we use a boundary integral method (BIM).
The method is well suited for problems with deforming boundaries such as RBC membranes \cite{pozrikidis01, zhao11}.
In order to solve for the particle velocity given an exerted force, a completed double layer boundary integral method (CDLBIM) \cite{power87, kohr04} has been combined with the classical BIM \cite{zhao12}.
The integral equations for the two-particle membrane systems read
\begin{equation} 
	\begin{split}
		v_{\beta}(\X) &= H_{\beta}(\X) \,  ,  \quad \X \in \Sm \, , \\
		\frac{1}{2} \phi_{\beta}(\X) + \sum_{\alpha=1}^{6} \varphi_{\beta}^{(\alpha)}(\X) \braket{\vect{\varphi}^{(\alpha)}, \vect{\phi}} &= H_{\beta}(\X) \, , \quad \X \in \Sp \, .
		\label{eq:CDL}
	\end{split}
\end{equation}
where $\Sm$ is the surface of the elastic membrane and $\Sp := S_{\mathrm{p}_{\gamma}} \cup S_{\mathrm{p}_{\lambda}}$ is the surface of the two spheres.
Here $\vect{v}$ denotes the velocity on the membrane whereas $\vect{\phi}$ represents the double layer density function on $\Sp$, related to the velocity of the particle~$\gamma$ via 
\begin{equation}
	{V_{\gamma}}_{\beta}(\X) = \sum_{\alpha=1}^{6} \varphi_{\beta}^{(\alpha)}(\X) \braket{\vect{\varphi}^{(\alpha)}, \vect{\phi}} \, , \quad \X \in S_{\mathrm{p}_{\gamma}} \, .
	\label{eq:CDL:Vel}
\end{equation}
where $\vect{\varphi}^{(\alpha)}$ are known functions \cite{kohr04}. 
The brackets stand for the inner product in the space of real functions whose domain is $S_{\mathrm{p}_{\gamma}}$, and the function $H_\beta$ is defined by
\begin{equation}
	H_\beta(\X) := -  (N_{\mathrm{m}} \Delta \vect{f})_\beta (\X) -  (K_\mathrm{p} \vect{\phi})_\beta(\X) + \mathcal{G}_{\beta\mu}^{(0)}(\X,\X_{\lambda_\mathrm{c}}) F_\mu   \, , \notag
\end{equation}
with $\X_{\lambda_\mathrm{c}}$ being the centroid of the sphere labeled $\lambda$ upon which the force is applied.
The single layer integral is defined as
\begin{equation}
	(N_{\mathrm{m}} \Delta \vect{f})_\beta (\X) := \int_{\Sm} \Delta f_\alpha(\Y) \mathcal{G}_{\alpha\beta}^{(0)}(\Y, \X) \, \Intd S(\Y) \notag
\end{equation}
and the double layer integral as
\begin{equation}
	(K_\mathrm{p} \vect{\phi})_\beta(\X) := \oint_{\Sp} \phi_\alpha(\Y) \mathcal{T}_{\alpha\beta\mu}^{(0)}(\Y,\X) n_\mu(\Y) \, \Intd S(\Y) \, . \notag
\end{equation}

Here, $\Delta \vect{f}$ is the traction jump, $\vect{n}$ denotes the outer normal vector at the particle surfaces and $\vect{F}$ is the force acting on the rigid particle.
The infinite-space Green's function is given by Eq.~\eqref{infiniteSpaceGreensFunction} and the corresponding Stresslet, defined as the symmetric part of the first moment of the force density, reads \cite{kim05}
\begin{equation}
 \mathcal{T}_{\alpha\beta\mu}^{(0)}(\Y,\X) = -\frac{3}{4\pi} \frac{s_\alpha s_\beta s_\mu}{s^5} \, , \notag
\end{equation}
with $\vect{s} := \Y - \X$ and $s := |\vect{s}|$.
The traction jump across the membrane $\Delta \vect{f}$ is an input for the equations, determined from the instantaneous deformation of the membrane.
In order to solve Eqs.~\eqref{eq:CDL} numerically, the membrane and particles' surfaces are discretized with flat triangles.
The resulting linear system of equations for the velocity $\vect{v}$ on the membrane and the density $\vect{\phi}$ on the rigid particles is solved iteratively by GMRES \cite{saad86}.
The velocity of each particle is determined from \eqref{eq:CDL:Vel}.
For further details concerning the algorithm and its implementation, we refer the reader to Ref. \cite{daddi16}.
Bending forces are computed using Method C from \cite{Guckenberger2016}.

In order to compute the particle self- and pair-mobilities numerically, a harmonic oscillating force $\vect{F}_{\lambda}(t) = \vect{A}_{\lambda} e^{i\omega_0 t}$ of amplitude $\vect{A}_{\lambda}$ and frequency $\omega_0$ is applied at the surface of the particle~$\lambda$.
After a brief transient time, both particles begin to oscillate at the same frequency as $\vect{V}_{\lambda} (t) = \vect{B}_{\lambda} e^{i(\omega_0 t + \delta_{\lambda})}$ and as $\vect{V}_{\gamma} (t) = \vect{B}_{\gamma} e^{i(\omega_0 t + \delta_{\gamma})}$.
The velocity amplitudes and phase shifts can accurately be obtained by a fitting procedure of the numerically recorded particle velocities.
For that, we use a nonlinear least-squares algorithm based on the trust region method \cite{conn00}.
Afterward, the $\alpha\beta$ component of the frequency-dependent complex self- and pair-mobilities can be calculated as
\begin{equation}
 \mu_{\alpha\beta}^{\lambda\lambda} = \frac{{B_{\lambda}}_{\alpha}}{{A_{\lambda}}_{\beta}} e^{i \delta_{\lambda}} \, , \quad
 \mu_{\alpha\beta}^{\gamma\lambda} = \frac{{B_{\gamma}}_{\alpha}}{{A_{\lambda}}_{\beta}} e^{i \delta_{\gamma}} \, .  \notag
\end{equation}


\section{Results}\label{sec:results}

For a single membrane, the corrections to the particle mobility can conveniently be split up into a correction due to shearing and area expansion together with a correction due to bending \cite{daddi16}. 
In the following, we denote by $\mu^{\gamma\gamma}_{\alpha\beta} = \mu^{\lambda\lambda}_{\alpha\beta}=\muS_{\alpha\beta}$ (``self'') the components of the self-mobility tensor,
and by $\mu^{\gamma\lambda} _{\alpha\beta} = \mu^{\lambda\gamma} _{\beta\alpha} = \muP_{\alpha\beta}$ (``pair'') the components of the pair-mobility tensor. 
Note that for $\alpha\neq\beta$,  $\muS_{\alpha\beta}=0$ and that $\muP_{\alpha\beta}=-\muP_{\beta\alpha}$. 

\subsection{Self-mobilities for finite-sized particles}

Mathematical expressions for the translational particle self-mobility corrections will be derived in terms of $\epsilon=a/z_0$.
The point-particle approximation presented in earlier work \cite{daddi16} represents the first order in the perturbation series, valid when the particle is far away from the membrane.

\subsubsection{Perpendicular to membrane}

The particle mobility perpendicular to the membrane is readily obtained after plugging the correction $\Delta \G_{zz}$ as defined by Eq.~\eqref{eqn:defDeltaG} to the normal-normal component of the Green's function from Eq.~\eqref{Gzz} into Eq.~\eqref{particleSelfMobility}. 
After computation, we find that the contribution due to shearing and bending can be expressed as
\begin{subequations}
\begin{align}
  \frac{\Delta \mu_{zz, \mathrm{S}}^{\mathrm{S}} }{\mu_0} &= e^{i\beta} \bigg( -\frac{9}{16}  \E_4 (i\beta) \epsilon 
 + \frac{3}{4} \E_5 (i\beta) \epsilon^3 \notag \\
 &- \frac{5}{16} \E_6 (i \beta) \epsilon^5 \bigg)  \, ,
 \label{deltaMuPerpShear} \\
  \frac{\Delta \mu_{zz, \mathrm{B}}^{\mathrm{S}} }{\mu_0} &= \epsilon f_1   + \epsilon^3 f_3   + \epsilon^5 f_5  \, , 
     \label{deltaMuPerpBending}
\end{align}
\end{subequations}
where the subscripts S and B stand for shearing and bending, respectively. The function $\E_n$ is the generalized exponential integral defined as $\E_n(x) := \int_1^\infty t^{-n} e^{-xt} \Intd t$ \cite{abramowitz72}.
Furthermore, $\beta := 6Bz_0 \eta \omega/\kS$ is a dimensionless frequency associated with the shearing resistance, whereas $B := 2/(1+C)$.
Moreover, $\betaB:=2z_0 (4\eta\omega/\kB)^{1/3}$ is a dimensionless number associated with bending. 
The functions $f_i$, with $i \in \{1,3,5\}$ are defined  by
 \begin{align}
  f_1  &= -\frac{15}{16} + \frac{3i \betaB}{8}
  \Bigg(
 \left( \frac{\betaB^2}{12}+\frac{i\betaB}{6} + \frac{1}{6}\right)\phi_+ \notag \\
 &+ \frac{\sqrt{3}}{6} (\betaB+i)\phi_{-}
 +  \left( \frac{\betaB^2}{12} - \frac{i\betaB}{3} - \frac{1}{3}\right) \psi
 \Bigg) \, , \notag \\
 f_3  &= \frac{5}{16} - \frac{\betaB^3}{48} \bigg( \left(\frac{\betaB}{4}+i \right)\phi_{+} + \frac{i\sqrt{3}\betaB}{4} \phi_{-} \notag \\
 &- \left( \frac{\betaB}{2}-i \right) \psi + \frac{3i}{2}
 \bigg) \, , \notag \\
 f_5  &= -\frac{1}{16} +\frac{\betaB^3}{384} \left( \frac{\betaB^2}{3} \left( \frac{\sqrt{3}}{2} \phi_{-} + \frac{i}{2} \phi_{+} - i\psi \right) + i \right) \, , \notag
  \end{align}
with
  \begin{align}
   \phi_{\pm} &:=  e^{-i \overline{z_{\mathrm{B}}}} \E_1 \left(-i\overline{z_{\mathrm{B}}} \right) \pm  e^{-i z_{\mathrm{B}}} \E_1 \left(-iz_{\mathrm{B}} \right) \, , \notag \\
  \psi &:= e^{-i\betaB}\E_1 (-i\betaB) \, , \notag
  \end{align}
where $z_{\mathrm{B}} := j\beta_\mathrm{B}$ and $j:=e^{2i\pi/3}$ being the principal cubic-root of unity. 
The bar designates complex conjugate. 

The total mobility correction is obtained by adding the individual contributions due to shearing and bending, as given by Eqs.~\eqref{deltaMuPerpShear} and \eqref{deltaMuPerpBending}.
In the vanishing frequency limit, the known result for a hard-wall \cite{swan10} is obtained:
\begin{equation}
 \lim_{\beta,\betaB \to 0} \frac{\Delta \mu_{zz}^{\mathrm{S}}}{\mu_0} = -\frac{9}{8} \epsilon + \frac{1}{2} \epsilon^3 -\frac{1}{8} \epsilon^5  \, .
 \label{hardWallPerp}
\end{equation}

The particle mobility near an elastic membrane is determined by membrane shearing and bending properties.
We therefore consider a typical case for which both effect manifests themselves equally.
For that purpose, we define a characteristic time scale for shearing as $\TS:=6z_0 \eta/\kS$ together with a characteristic time scale for bending as $\TB := 4\eta z_0^3/\kB$ \cite{daddi16}. Then we take $z_0^2 \kS / \kB = 3/2$ such that the two time scales are equal and can be denoted by $\TS=\TB=:T$.
In this case, the two dimensionless numbers $\beta$ and $\betaB$ are related by $\betaB = 2 (\beta/B)^{1/3}$. 
The situation for a membrane with the typical parameters of a red blood cell is qualitatively similar as shown in the Supporting Information.
\footnote{See Supplemental Material at [URL will be inserted by publisher] for the frequency-dependent mobilities where typical values for the RBC parameters are used.}

In Fig.~\ref{mobiCorrSelf} $a)$, we show the particle scaled self-mobility corrections versus the scaled frequency $\beta$, as stated by Eqs.~\eqref{deltaMuPerpShear} and \eqref{deltaMuPerpBending}.
The particle is set a distance $z_0 = 2a$ above the membrane. 
We observe that the real part is a monotonically increasing function with respect to frequency while the imaginary part exhibits a bell-shaped dependence on frequency centered around $\beta \sim 1$.
In the limit of infinite frequencies, both the real and imaginary parts of the self-mobility corrections vanish, and thus one recovers the bulk behavior.
For the perpendicular motion we observe that the particle mobility correction is primarily determined by the bending part.

A very good agreement is obtained between the analytical predictions and the numerical simulations over the whole range of frequencies. Additionally, we assess the accuracy of the point-particle approximation employed in earlier work \cite{daddi16}, in which only the first order correction term in the perturbation parameter $\epsilon$ was considered. While this approximation slightly underestimates particle mobilities, it nevertheless leads to a  surprisingly good prediction, even though  the particle is set only one diameter above the membrane.


\subsubsection{Parallel to membrane}

We proceed in a similar way for the motion parallel to the membrane. 
By plugging the correction $\Delta \G_{xx}$ from the Green's function in Eq.~\eqref{Gxx} into Eq.~\eqref{particleSelfMobility} we find
\begin{subequations}
\begin{align}
	\frac{\Delta \mu_{xx,\mathrm{S}}^{\mathrm{S}} }{\mu_0} &= e^{i\beta} \bigg( -\frac{3}{32} \bigg(3\E_4(i\beta)-4\E_3(i\beta)+2\E_2(i\beta) \notag \\
	&+ 4 e^{iC\beta} \E_2(i(1+C)\beta) \bigg) \epsilon \label{deltaMuParaShear} \\
	&+ \frac{3}{16} \left( 2\E_5(i\beta) - \E_4 (i\beta) \right) \epsilon^3 
	- \frac{5}{32} \E_6 (i\beta) \epsilon^5 \bigg)  \, ,
	\notag \\
      \frac{\Delta \mu_{xx,\mathrm{B}}^{\mathrm{S}} }{\mu_0} &= \epsilon g_1   + \epsilon^3 g_3   + \epsilon^5 g_5   \, , 
      \label{deltaMuParaBending}
\end{align}
\end{subequations}
where we defined
 \begin{align}
  g_1  &= -\frac{3}{32} + \frac{i \betaB^3}{64} (\phi_{+} + \psi) \, , \notag \\
  g_3  &= \frac{3}{32} + \frac{\betaB^3}{64} \left( -i + \frac{\betaB}{3} \left( \psi - \frac{1}{2}\phi_{+} - \frac{i\sqrt{3}}{2} \phi_{-} \right) \right) \, , \notag \\
  g_5  &= -\frac{1}{32} +\frac{\betaB^3}{768} \left(i + \frac{\betaB^2}{3} \left( \frac{i}{2}\phi_{+} + \frac{\sqrt{3}}{2} \phi_{-} - i\psi \right)  \right) \ . \notag
 \end{align}

The well-known hard-wall limit, as first calculated by \Faxen \cite{faxen22, swan10}, is recovered by considering the vanishing frequency limit:
\begin{equation}
 \lim_{\beta,\betaB \to 0} \frac{\Delta \mu_{xx}^{\mathrm{S}}}{\mu_0} = -\frac{9}{16} \epsilon + \frac{1}{8} \epsilon^3 -\frac{1}{16} \epsilon^5  \, .
  \label{hardWallPara}
\end{equation}

\begin{figure}
  \begin{center}
     \includegraphics[scale = 0.42]{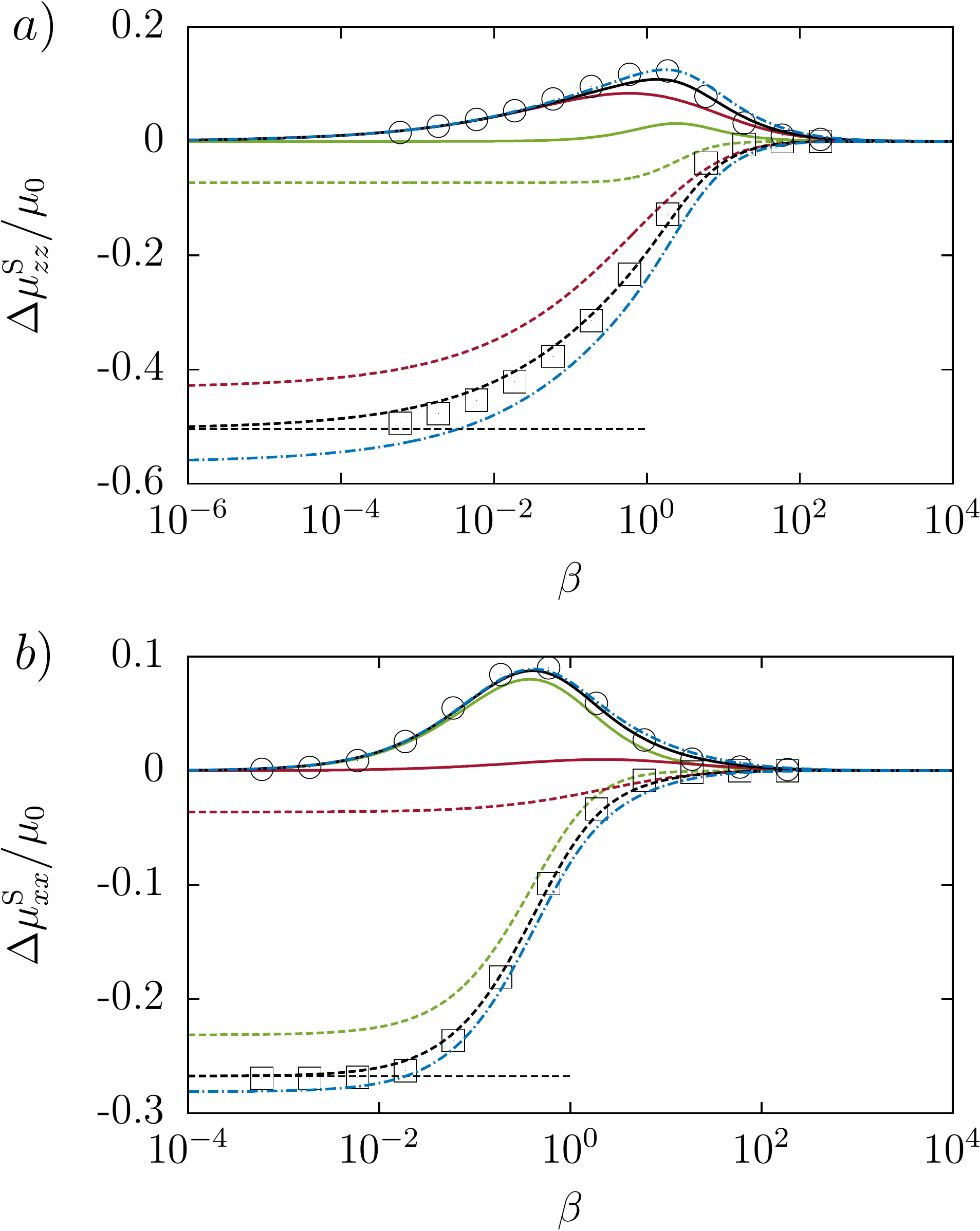}
     \caption{ (Color online) The scaled frequency-dependent self-mobility correction versus the scaled frequency for the motion perpendicular $(a)$ and parallel $(b)$ to the membrane.
     The particle is located at $z_0 = 2a$. We take $z_0^2 \kS / \kB = 3/2$ and $C=1$ in the Skalak model.
     The analytical predictions are shown as dashed lines for the real part, and as solid lines for the imaginary part.
     Symbols refer to BIM simulations. 
     The shearing and bending contributions are shown in green and red respectively.
     The dotted-dashed line in blue corresponds to the first order correction in the particle self-mobility, as previously determined in Ref. \cite{daddi16}.
     Horizontal dashed lines represent the mobility corrections near a hard-wall as given by Eqs.~\eqref{hardWallPerp} and \eqref{hardWallPara}. 
     }
     \label{mobiCorrSelf}
  \end{center}
\end{figure}

The mobility corrections in the parallel direction are shown in Fig.~\ref{mobiCorrSelf} $b)$. We observe that the total correction is mainly determined by the shearing part in contrast to the perpendicular case where bending dominates.


\subsection{Pair-mobilities for finite-sized particles}

In the following, expressions for the pair-mobility corrections in terms of a power series in $\sigma=a/h$ will be provided.
To start, let us first recall the particle pair-mobilities in an unbounded geometry.
By applying Eq.~\eqref{particlePairMobility} to the infinite space Green's function Eq.~\eqref{infiniteSpaceGreensFunction}, the bulk pair-mobilities for the motion perpendicular to and along the line of centers read \cite[p. 190]{kim05}
\begin{equation}
 \frac{ \mu_{zz}^{\mathrm{P}}}{\mu_0} = \frac{3}{4} \sigma + \frac{1}{2} \sigma^3  \, , \quad 
 \frac{ \mu_{xx}^{\mathrm{P}}}{\mu_0} = \frac{3}{2} \sigma - \sigma^3  \, ,
\end{equation}
and are commonly denominated the Rotne-Prager tensor \cite{rotne69, ermak78}. 
Note that the terms with $\sigma^5$ vanish for the bulk mobilities when considering only the first reflection as is done here.
The axial symmetry along the line connecting the two spheres in bulk requires that $\mu_{yy}^{\mathrm{P}} = \mu_{zz}^{\mathrm{P}}$ and that the off-diagonal components of the mobility tensor are zero.
Physically, the parameter $\sigma$ only takes values between 0 and $1/2$ as overlap between the two particles is not allowed.
In this interval, the pair-mobility perpendicular to the line of centers $\mu_{zz}^{\mathrm{P}}$ is always lower than the pair-mobility $\mu_{xx}^{\mathrm{P}}$, since it is easier to move the fluid aside than to push it into or to squeeze it out of the gap between the two particles.


\begin{figure*}
  \begin{center}
     \includegraphics[scale = 0.8]{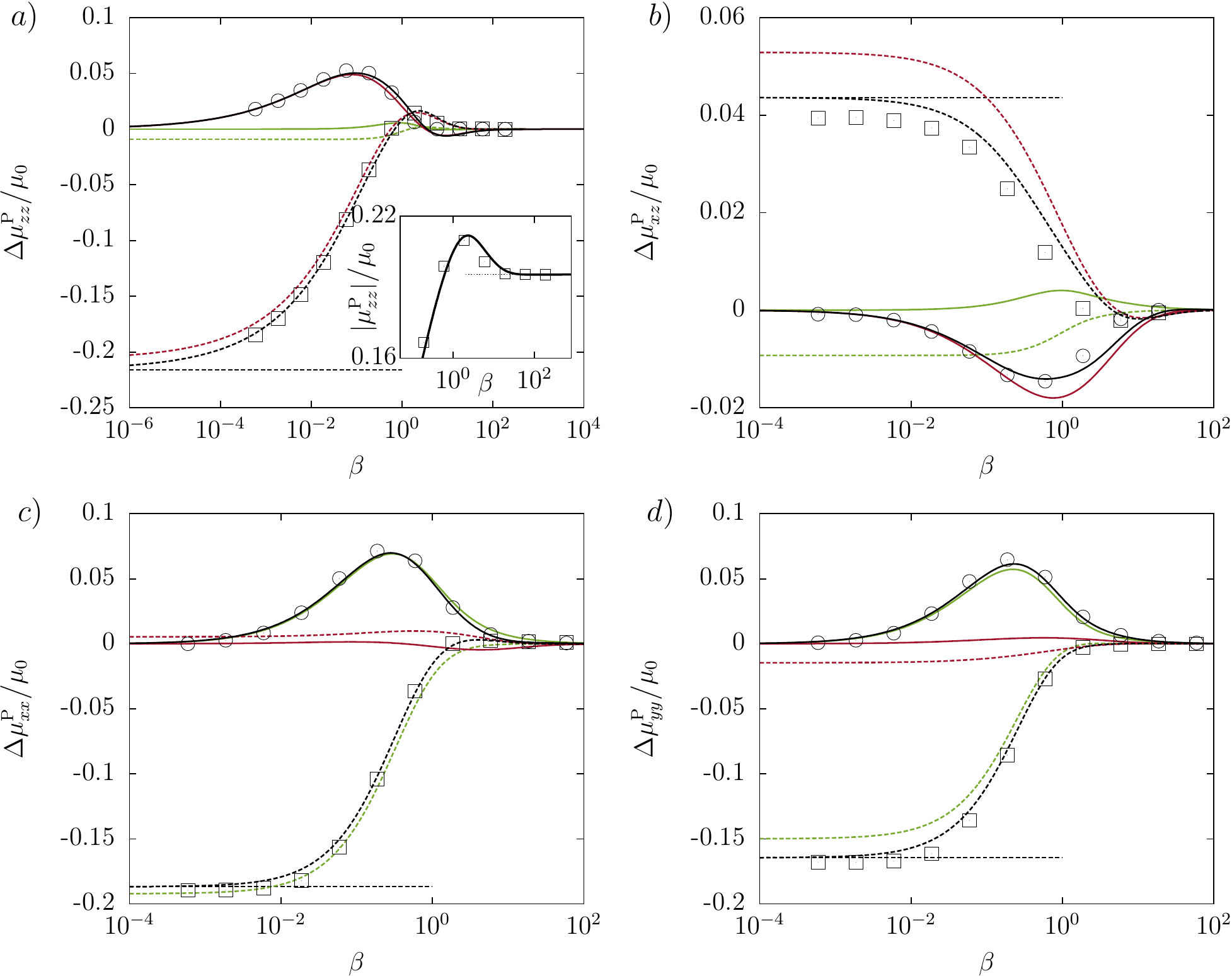}
     \caption{ (Color online) The scaled pair-mobility corrections versus the scaled frequency $\beta$.
     The two particles are located above the membrane at $z_0 = 2a$ with a distance $h=4a$. 
     The real and imaginary parts of the mobility correction are shown as dashed and solid lines, respectively.
     The shearing and bending related parts are shown in green and red, respectively. The hard-wall limits are shown as horizontal dashed lines.
     The inset in $a)$ shows that the amplitude of the total pair-mobility component $zz$ exceeds its bulk value (dotted line) in a small frequency range around $\beta\sim 1$. }
     \label{mobiCorr}
  \end{center}
\end{figure*}

Consider next the pair-mobilities near an elastic membrane.
By applying  Eq.~\eqref{particlePairMobility} to Eqs.~\eqref{Gzz} through \eqref{Gxz}, we find that the corrections to the pair-mobilities can conveniently be expressed in terms of the following convergent integrals,
\begin{subequations}
\begin{align}
 \frac{\Delta \mu_{zz}^{\mathrm{P}} }{\mu_0} &= \int_{0}^{\infty} - \frac{i \sigma u^3}{3\xi^{5/2}} \left( \frac{ \Lambda^2}{ 2iu-\beta}
 + \frac{4 \Gamma_{-}^2 }{8iu^3-\betaB^3} \right) \notag \\
 &\times \chi_0 e^{-2u} \Intd u \, , \label{muZZ_Elastic} \\
 \frac{\Delta \mu_{xx}^{\mathrm{P}}  }{\mu_0} &=  \int_{0}^{\infty} \bigg( \frac{i\sigma}{6\xi^{5/2}} \left( \frac{\Gamma_{+}^2}{2iu-\beta} + \frac{4u^4 \Lambda^2}{8i u^3-\betaB^3} \right) \notag \\
 &\times \left( \xi^{1/2} \chi_1 -2u \chi_0 \right) 
  -\frac{3\sigma B}{2} \frac{ \chi_1}{B u+i\beta} \bigg) e^{-2u} \Intd u \, , \label{muXX_Elastic} \\
  \frac{\Delta \mu_{yy}^{\mathrm{P}} }{\mu_0} &=  \int_{0}^{\infty} \bigg( -\frac{i\sigma}{6\xi^{2}} \left( \frac{\Gamma_{+}^2}{2iu-\beta} + \frac{4u^4 \Lambda^2}{8i u^3-\betaB^3} \right) 
  \chi_1 \notag \\
  &+\frac{3\sigma B}{2 \xi^{1/2}} \frac{ \xi^{1/2} \chi_1 -2u \chi_0}{B u+i\beta} \bigg) e^{-2u} \Intd u \, , \label{muYY_Elastic} \\
  \frac{\Delta \mu_{xz}^{\mathrm{P}} }{\mu_0} &= \int_{0}^{\infty} \frac{i\sigma u^2}{3\xi^{5/2}} \Lambda \left( \frac{\Gamma_{+}}{2iu-\beta}+\frac{4u^2 \Gamma_{-}}{8iu^3-\betaB^3} \right) \notag \\
  &\times \chi_1 e^{-2u} \Intd u \, , \label{muXZ_Elastic}
\end{align}
\end{subequations}
where $\xi := 4z_0^2/h^2 =  4\sigma^2/\epsilon^2$ 
and
\begin{eqnarray}
\Lambda &:=& 4\sigma^2 u - 3\xi \, ,\notag\\
\Gamma_{\pm} &:=& 4\sigma^2 u^2-3u\xi \pm 3\xi \,,\notag\\
\chi_n &:=& J_n \left( \frac{2u}{\xi^{1/2}} \right) \, . \notag
\end{eqnarray}

The terms involving $\beta$ and $\betaB$ in Eqs.~\eqref{muZZ_Elastic} through \eqref{muXZ_Elastic} are the contributions coming from shearing and bending, respectively. Due to symmetry, $\mu_{\alpha y}^{\mathrm{P}} = 0$ for $\alpha \in \{x,z\}$.

For future reference, we note that each component of the frequency-dependent particle self- and pair-mobility tensor can conveniently be cast in the form
\begin{equation}
 \frac{\mu (\omega)}{\mu_0} = b + \int_{0}^{\infty}\frac{\varphi_1 (u)}{\varphi_2 (u) + i\omega T } \Intd u \, , 
 \label{eqn:muGeneral}
\end{equation}
where indices and superscripts have been omitted.
Here $b$ denotes the scaled bulk mobility (cf.~Eq.~\eqref{eqn:defMobility}), and the integral term represents either shearing or bending related parts in the mobility correction.
Note that $\varphi_1$ and $\varphi_2$ are real functions which do not depend on frequency. 
Moreover, $\varphi_2 (u)=2u/B$ or $\varphi_2 (u)=u$ for the shearing related parts and $\varphi_2 (u) = u^3$ for bending such that $\varphi_2 (u) \ge 0$, $\forall u \in [0,\infty)$.

In the vanishing frequency limit, i.e.\ for $\beta,\betaB$ both taken to zero we recover the pair-mobilities near a hard-wall with stick boundary conditions, namely 
\begin{subequations}
\begin{align}
 \frac{\Delta \mu_{zz}^{\mathrm{P}}}{\mu_0} &= -\frac{3}{4}\frac{3\xi^2+\frac{5}{2}\xi+1}{(1+\xi)^{5/2}} \sigma 
				       +\frac{4\xi^2 - 4\xi - \frac{1}{2}}{(1+\xi)^{7/2}} \sigma^3 \notag \\
				       &-\frac{4\xi^2 - 12\xi + \frac{3}{2}}{(1+\xi)^{9/2}} \sigma^5 \, , \label{muZZ} \\
 \frac{\Delta \mu_{xx}^{\mathrm{P}}}{\mu_0} &= -\frac{3}{2}\frac{1+\xi+\frac{3}{4} \xi^2}{(1+\xi)^{5/2}} \sigma 
				       +\frac{\xi^2 -\frac{11}{2}\xi + 1}{(1+\xi)^{7/2}} \sigma^3 \notag \\
				       &-\frac{2\xi^2-\frac{27}{2}\xi + 2}{(1+\xi)^{9/2}} \sigma^5 \, , \label{muXX} \\
 \frac{\Delta \mu_{yy}^{\mathrm{P}}}{\mu_0} &= -\frac{3}{4}\frac{1+\frac{3}{2} \xi}{(1+\xi)^{3/2}} \sigma
				       +\frac{\xi - \frac{1}{2}}{(1+\xi)^{5/2}} \sigma^3
				       -\frac{2\xi - \frac{1}{2}}{(1+\xi)^{7/2}} \sigma^5 \, , \label{muYY} \\
 \frac{\Delta \mu_{xz}^{\mathrm{P}}}{\mu_0} &= \frac{9}{8} \frac{\xi^{3/2}}{(1+\xi)^{5/2}} \sigma
				       -\frac{3}{2} \frac{(4\xi-1)\xi^{1/2}}{(1+\xi)^{7/2}} \sigma^3 \notag \\
				       &+\frac{5}{2} \frac{(4\xi-3)\xi^{1/2}}{(1+\xi)^{9/2}} \sigma^5 \, , \label{muXZ}
\end{align}
\end{subequations}
in agreement with the results by Swan and Brady \cite{swan07}.


In Fig.~\ref{mobiCorr} we plot the particle pair-mobilities as given by Eqs.~\eqref{muZZ_Elastic} through \eqref{muXZ_Elastic} as functions of the dimensionless frequency $\beta$ for $h=4a$.
We observe that the real and imaginary parts have basically the same evolution as the self-mobilities. 
Nevertheless, two qualitatively different effects are apparent from Fig.~\ref{mobiCorr}:
First, 
the amplitude of the normal-normal pair-mobility $\left|\muP_{zz}\right|$ in a small frequency range even exceeds its bulk value.
This enhanced mobility results in a short-lasting superdiffusive behavior as will be described in Sec.~\ref{sec:diffusion}.

Secondly, for the components $xx$ and $xz$ in Fig.~\ref{mobiCorr} we find that, unlike the self-mobilities, shearing and bending may have opposite contributions to the total pair-mobilities. For the $xz$ component this implies the interesting behavior that hydrodynamic interactions can be either attractive or repulsive depending on the membrane properties. This will be investigated in more detail in the next subsection.


\subsection{Perpendicular steady motion}

\begin{figure}
  \begin{center}
     \includegraphics[scale = 0.35]{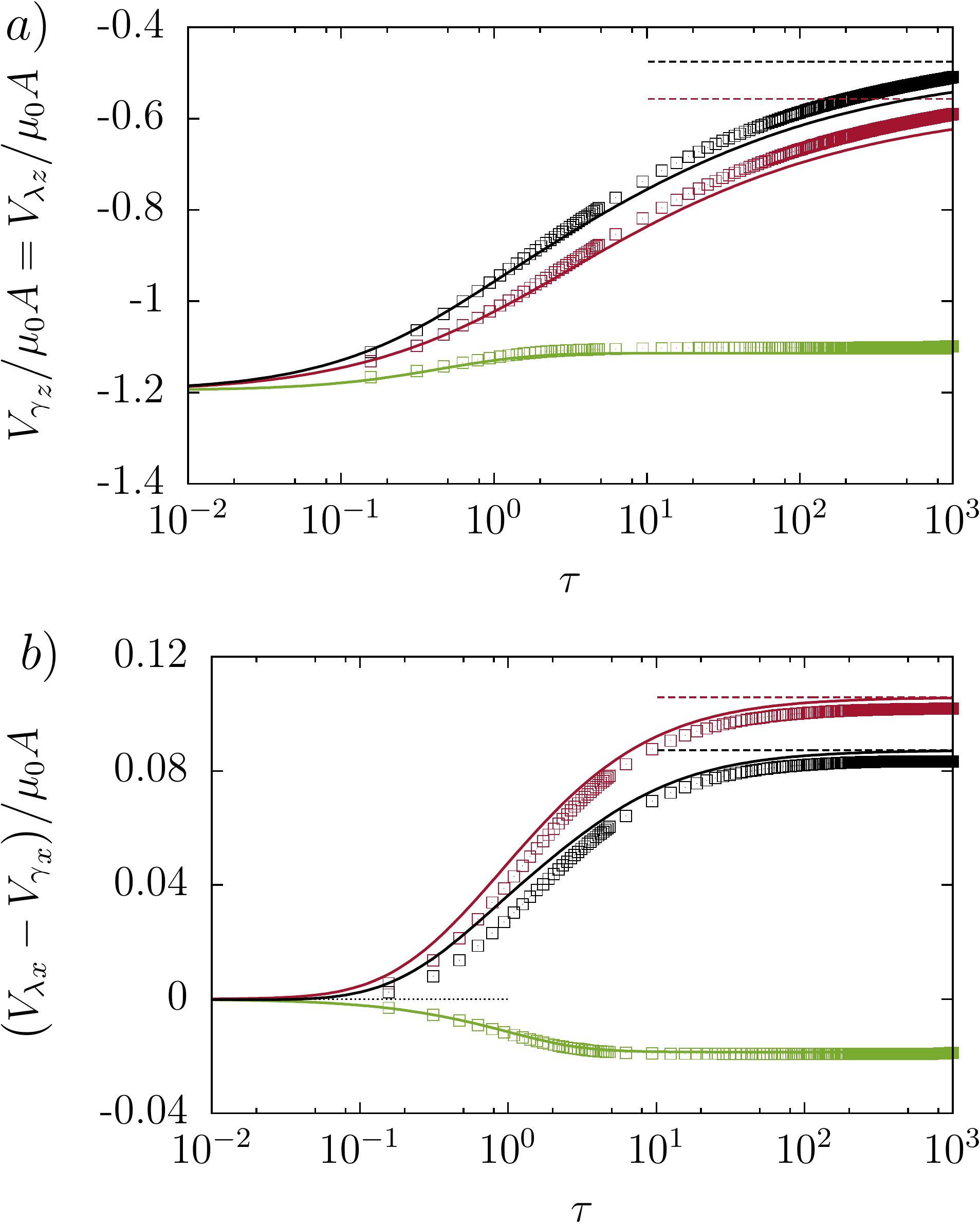}
     \caption{ (Color online) The scaled particle velocities perpendicular to the membrane ($a$) and relative to each other ($b$) versus the scaled time for a constant force acting downward on both particles near a membrane endowed with only shearing (green),  only bending (red) or both rigidities (black).
     Solid lines are the analytical predictions as given by Eq.~\eqref{velocityStepForce}, symbols are obtained by boundary-integral simulations.
     Horizontal dotted and dashed lines stand for the bulk and vanishing frequency limits respectively. 
} 
     \label{attractionRepulsionGeschDown}
  \end{center}
\end{figure}

A situation in which hydrodynamic interactions are particularly relevant is the steady approach of two particles towards an interface, such as e.g.\ drug molecules approaching a cell membrane, reactant species approaching a catalyst interface, charged colloids being attracted by an oppositely charged membrane, etc. For hard walls, it is known that hydrodynamic interactions in this case are repulsive \cite{dufresne00, squires00, swan07} leading to the dispersion of particles on the surface.
Near elastic membranes, the different signs of the bending and shear contributions to the pair-mobility in Fig.~\ref{mobiCorr}~$b)$ point to a much more complex scenario including the possibility of particle attraction.

The physical situation of two particles being initially located at $z=z_0$ and suddenly set into motion towards the interface is described by a Heaviside step function force $\vect{F} (t) = \vect{A} \theta(t)$. Its Fourier transform to the frequency domain reads \cite{bracewell99}
\begin{equation}
 \vect{F} (\omega) = \left( \pi \delta (\omega) -\frac{i}{\omega} \right) \vect{A} \, . \notag
\end{equation}
Using the general form of Eq.~\eqref{eqn:muGeneral}, the scaled particle velocity in the temporal domain is then given by
\begin{equation}
 \frac{V(\tau)}{\mu_0 A} = \left( b + \int_{0}^{\infty} \frac{\varphi_1(u)}{\varphi_2(u)} \left( 1-e^{-\varphi_2(u) \tau} \right) \Intd u  \right) \theta (\tau) \, ,
 \label{velocityStepForce}
\end{equation}
where $\tau := t/T$ is a dimensionless time.
At larger times, the exponential in Eq.~\eqref{velocityStepForce} can be neglected compared to one.
In this way, we recover the steady velocity near a hard-wall.

In corresponding BIM simulations, a constant force of small amplitude towards the wall is applied on both particles in order to retain the system symmetry.
At the end of the simulations, the vertical position of the particles changes by about 8 \% compared to their initial positions $z_0$.

In Fig.~\ref{attractionRepulsionGeschDown}~$a)$ we show the time dependence of the vertical velocity which at first increases and then approaches its steady-state value. Figure~\ref{attractionRepulsionGeschDown}~$b)$ shows the relative velocity between the two particles: clearly, the motion is attractive for a membrane with negligible bending resistance (such as a typical artificial capsule) which is the opposite of the behavior near a membrane with only bending resistance (such as a vesicle) or a hard wall.

\begin{figure}
  \includegraphics[scale = 5]{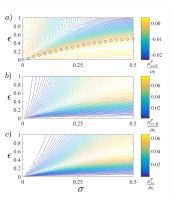}
  \caption{(Color online) Contour diagram $(\epsilon,\sigma)$ of the shearing $(a)$ and bending $(b)$ contribution in the vanishing frequency limit in the $xz$ pair-mobility
   as stated by Eq.~\eqref{muXZ_S} for shearing and by Eq.~\eqref{muXZ_B} for bending.
   $c)$ is the same contour near a hard-wall as given by Eq.~\eqref{muXZ}.
   The perturbation solution given by Eq.~\eqref{epsilon_XZ_line} is presented as circles in $(a)$. Contrary to a membrane with bending resistance and to a hard wall, the mobility changes sign near a membrane with shearing resistance. This sign change directly reflects the physically observable situation as the bulk contribution for the $xz$ pair-mobility is zero.
   }
 \label{ContoursXZ}
\end{figure}

In order to illustrate more clearly for which wall and particle distances a repulsion/attraction is expected we show in Fig.~\ref{ContoursXZ} the pair-mobility correction for the shear $\Delta\muP_{xz, \mathrm{S}}$ and bending $\Delta\muP_{xz, \mathrm{B}}$ contributions in the $(\epsilon,\sigma)$ plane. 
To reduce the parameter space and to bring out the considered effects most clearly, we consider the idealized limit $\omega\to 0$. In this limit, the contributions become independent of the elastic moduli since $\omega\to 0$ directly implies that $\beta,\beta_B\to 0$ meaning that even infinitesimally small shearing and bending resistances would make the membrane behave identical to the hard wall. This unphysical behavior is remedied in a realistic situation where a small bending resistance will lead to a correspondingly large time scale $\TB$ and thus to a long-lived transient regime as given by Eq.~\eqref{velocityStepForce} and shown in the Supporting Information. Therefore, the contours shown in Fig.~\ref{ContoursXZ} faithfully represent the behavior of membranes with small bending (Fig.~\ref{ContoursXZ}~$a)$) or small shear (Fig.~\ref{ContoursXZ}~$b)$) resistance. The corresponding equations can be found in Appendix~\ref{appendix:vanishingFrequencyLimits}.

By equating Eq.~\eqref{muXZ_S} to zero and solving the resulting equation perturbatively, the threshold lines where the shearing contribution changes sign are given up to fifth order in $\sigma$ by
\begin{equation}
 \ETH =  \sqrt{2}  \left( \sigma - \frac{4}{3} \sigma^3 + \frac{17}{27} \sigma^5  \right) + \bigO (\sigma^7) \, . \label{epsilon_XZ_line}
\end{equation}
Eq.~\eqref{epsilon_XZ_line} is shown as circles in Fig.~\ref{ContoursXZ}.
The bending contribution in Fig.~\ref{ContoursXZ}~$b)$ always has a positive sign corresponding to a repulsive interaction similar as the hard wall.

Similar changes in sign are observed for $\Delta\muP_{zz, \mathrm{S}}$ for shear and $\Delta\muP_{xx, \mathrm{B}}$ for bending.
The corresponding contours are given in the Supporting Information. 
Their physical relevance, however, is less important than for $\Delta\muP_{xz}$ shown in Fig.~\ref{ContoursXZ} as the effects may be overshadowed by the bulk values of the mobilities (which is zero only for $\muP_{xz}$).


\section{Diffusion} \label{sec:diffusion}

The diffusive dynamics of a pair of Brownian particles is governed by the generalized Langevin equation written for each velocity component of particle $\gamma$ as \cite{kubo66}
\begin{equation}
 \begin{split}
  m \frac{\Intd {V_{\gamma}}_{\alpha}}{\Intd t} &= -\int_{-\infty}^{t} \zeta^{\gamma\gamma}_{\alpha\beta}(t-t') {V_{\gamma}}_{\beta} (t') \Intd t'  \\
                                      &-\int_{-\infty}^{t} \zeta^{\gamma\lambda}_{\alpha\beta}(t-t') {V_{\lambda}}_{\beta} (t') \Intd t'
                                       + {F_{\gamma}}_{\alpha} (t) \, .
 \label{generalizedLangevinEqn1}
 \end{split}
\end{equation}

A similar equation can be written for the velocity components of the other particle $\lambda$.
Here, $m$ denotes the particles' mass, $\zeta^{\gamma\lambda}_{\alpha\beta}(t)$ stands for the time-dependent two-particle friction retardation tensor (expressed in kg/s$^2$) and ${F_{\gamma}}_{\alpha}$ is a random force which is zero on average.
By evaluating the Fourier transform of both members in Eq.~\eqref{generalizedLangevinEqn1} and using the change of variables $w=t-t'$ together with  the shift property in the time domain of Fourier transforms we get
\begin{equation}
\begin{split}
  &im\omega {V_\gamma}_\alpha (\omega) + \zeta^{\gamma\gamma}_{\alpha\beta}[\omega]  {V_{\gamma}}_{\beta}(\omega) 
              + \zeta^{\gamma\lambda}_{\alpha\beta} [\omega] {V_{\lambda}}_{\beta}(\omega) = {F_{\gamma}}_{\alpha} (\omega) \, , \label{eq:GLE} \\
\end{split}
\end{equation}
where $\zeta^{\gamma\lambda}_{\alpha\beta} [\omega]$ and $\zeta^{\gamma\gamma}_{\alpha\beta} [\omega]$ are the Fourier-Laplace transforms of the retardation function defined as
\begin{equation}
 \zeta^{\gamma\lambda}_{\alpha\beta}[\omega] := \int_{0}^{\infty} \zeta^{\gamma\lambda}_{\alpha\beta}(t) e^{-i \omega t} \Intd t \, , \label{eq:FourierLaplcetransform}
\end{equation}
and analogously for $\zeta^{\gamma\gamma}_{\alpha\beta} [\omega]$.

In the following, we shall consider the overdamped regime for which the particles are massless $(m=0)$.
Solving Eq.~\eqref{eq:GLE} for the particle velocities and equating with the definition of the mobilities,
\begin{subequations}
\begin{align}
 {V_\gamma}_\alpha (\omega) &= \mu_{\alpha\beta}^{\gamma\gamma} (\omega) {F_\gamma}_\beta (\omega) + \mu_{\alpha\beta}^{\gamma\lambda}(\omega) {F_\lambda}_\beta (\omega) \, , \label{V1} \\
 {V_\lambda}_\alpha (\omega) &= \mu_{\alpha\beta}^{\lambda\lambda}(\omega) {F_\lambda}_\beta (\omega) +\mu_{\alpha\beta}^{\lambda\gamma} (\omega) {F_\gamma}_\beta (\omega) \,, \label{V2}
\end{align}
\end{subequations}
leads to expressions of the mobilities in terms of the friction coefficients:
\begin{subequations}
 \begin{align} 
  \muS_{xx}(\omega) &= \frac{\zetaXXS \zetaZZP} {({\zetaXXS}^2-{\zetaXXP}^2) \zetaZZP - \zetaXXP{\zetaXZP}^2} \, , \notag \\
  \muP_{xx}(\omega) &= -\frac{\zetaXXP \zetaZZP} {({\zetaXXS}^2-{\zetaXXP}^2) \zetaZZP - \zetaXXP{\zetaXZP}^2} \, , \notag \\
  \muS_{yy}(\omega) &= \frac{ \zetaYYS}{ {\zetaYYS}^2 - {\zetaYYP} ^2} \, , \notag \\
  \muP_{yy}(\omega) &= -\frac{\zetaYYP}{ {\zetaYYS} ^2 - {\zetaYYP} ^2} \, . \notag \\
  \muS_{zz}(\omega) &= \frac{\zetaXXS \zetaZZS} {({\zetaZZS}^2-{\zetaZZP}^2) \zetaXXS - \zetaZZS{\zetaXZP}^2} \, , \notag \\
  \muP_{zz}(\omega) &= -\frac{\zetaXXS \zetaZZP} {({\zetaZZS}^2-{\zetaZZP}^2) \zetaXXS - \zetaZZS{\zetaXZP}^2} \, , \notag \\
  \muP_{xz}(\omega) &= -\frac{\zetaZZS \zetaXZP} {({\zetaZZS}^2-{\zetaZZP}^2) \zetaXXS - \zetaZZS{\zetaXZP}^2} \, , \notag
 \end{align}
\end{subequations}
where the brackets [~] are dropped out for the sake of clarity. 
Similar as for the mobilities, the self- and pair components of the retardation function are denoted by $\zeta_{\alpha\beta}^{\gamma\gamma} = \zeta_{\alpha\beta}^{\lambda\lambda} = \zeta_{\alpha\beta}^{\mathrm{S}}$
and $\zeta_{\alpha\beta}^{\gamma\lambda}  = \zeta_{\beta\alpha}^{\lambda\gamma} = \zeta_{\alpha\beta}^{\mathrm{P}}$, respectively.
Note that $\zetaXXP \zetaZZS = \zetaXXS \zetaZZP$ so that $\muS_{xz} = 0$ as required by symmetry.


According to the fluctuation-dissipation theorem, the frictional and random forces are related via \cite{kubo85}[p. 33]\cite{Kheifets_2014}
\begin{equation}
\langle F_{\gamma} (\omega)  \overline{F_{\lambda} (\omega')} \rangle = \kBolt T  \left( \zeta_{\alpha\beta}^{\gamma\lambda} [\omega] + \overline{\zeta_{\alpha\beta}^{\gamma\lambda} [\omega]} \right)\delta(\omega-\omega') \, ,
 \label{powerSpectrumForce}
\end{equation}
and analogously for the $\gamma\gamma$ component, where $\kBolt$ is the Boltzmann constant and $T$ is the absolute temperature of the system.
\footnote{In Ref. \cite{kubo85}, a factor $2\pi$ appears in the denominator of Eq.~\eqref{powerSpectrumForce} in contrast to the present work, 
as they consider the factor $2\pi$ in the forward Fourier transform (left-hand side) and we consider it in the inverse transform while the Laplace transform (right-hand side) is defined identically.}

Multiplying Eq.~\eqref{V1} by its complex conjugate, taking the ensemble average and using Eq.~\eqref{powerSpectrumForce}, it can be shown that the velocity power spectrum obeys the relation
\begin{equation}
 {\mathcal{P}_{\mathrm{V}}}_{\alpha\beta}^{\mathrm{S}} (\omega) = \kBolt T \left( \muS_{\alpha\beta} (\omega) + \overline{\muS_{\alpha\beta} (\omega)} \right) \, . \label{eq:phiVS}
\end{equation}
Next, by considering both Eqs.~\eqref{V1} and \eqref{V2} together with Eq.~\eqref{powerSpectrumForce} we find in a similar fashion
\begin{equation}
 {\mathcal{P}_{\mathrm{V}}}_{\alpha\beta}^{\mathrm{P}} (\omega) = \kBolt T \left( \muP_{\alpha\beta} (\omega) + \overline{\muP_{\alpha\beta} (\omega)} \right) \, . \label{eq:phiVD}
\end{equation}

According to the Wiener-Khinchin-Einstein theorem, the velocity auto/cross-correlation functions can directly be obtained from the temporal inverse Fourier transform as \cite{kubo85} 
\begin{equation}
 {\phi}_{\alpha\beta}^{\gamma\lambda} (t)  = \frac{\kBolt T}{2\pi} \int_{-\infty}^{\infty} 
                                  \left( \mu^{\gamma\lambda}_{\alpha\beta} (\omega) + \overline{ \mu^{\gamma\lambda}_{\alpha\beta} (\omega)}  \right) e^{i\omega t} \Intd \omega \, .
                                  \label{eq:correlationFct}
\end{equation}

It can be shown using the residue theorem \cite[p. 34]{kubo85} that the integral over the second term in Eq.~\eqref{eq:correlationFct} vanishes if the mobility is an analytic function for $\operatorname{Im} (\omega)<0$. The present mobilities all fulfill this condition as can be seen by their general form in Eq.~\eqref{eqn:muGeneral}.


Most commonly, diffusion is studied in terms of the mean-square displacement (MSD) which can be calculated from the correlation function as \cite[p. 37]{kubo85}
\begin{equation}
\langle \Delta {r_{\gamma}}_{\alpha} (t) \Delta {r_{\lambda}}_{\beta} (t) \rangle = 2 \int_{0}^{t} (t-s) {\phi}_{\alpha\beta}^{\gamma\lambda} (s) \Intd s \, ,
\label{eq:MSD}
\end{equation}
where $\Delta {r_{\gamma}}_{\alpha}$ denotes the displacement of the particle~$\gamma$ in the direction $\alpha$.
Furthermore, we define the time-dependent pair-diffusion tensor as
\begin{equation}
 D_{\alpha\beta}^{\gamma\lambda}  (t) := \frac{\langle \Delta {r_{\gamma}}_{\alpha} (t) \Delta {r_{\lambda}}_{\beta} (t) \rangle}{2t} \, . \label{pairDiff}
\end{equation}
Analogous relations to Eqs.~\eqref{eq:correlationFct}-\eqref{pairDiff} hold for the $\gamma\gamma$ component.
We now consider the collective motions of the center of mass $\boldsymbol{\rho} := \R_{\lambda} + \R_{\gamma}$ as well as the relative motion $\vect{h} := \R_{\lambda} - \R_{\gamma}$ with the corresponding diagonal pair-diffusion tensor
\begin{equation}
 D_{\alpha \alpha}^{C, R} =  2 \left(D_{\alpha\alpha}^{\mathrm{S}} \pm D_{\alpha\alpha}^{\mathrm{P}}   \right) \, , 
 \label{collectiveRelativeModeDef}
\end{equation}
where the positive sign applies for the collective mode of motion and the negative sign to the relative mode. 
In the absence of the membrane, Eqs.~\eqref{collectiveRelativeModeDef} reduces to the generalization of the Einstein relation as calculated by Batchelor \cite{batchelor76} for the relative mode, namely
\begin{equation}
 \frac{D_{zz}^{\mathrm{R}}}{2D_0} = 1 - \frac{3}{4} \sigma - \frac{\sigma^3}{2}  \, , \quad 
 \frac{D_{xx}^{\mathrm{R}}}{2D_0} = 1 - \frac{3}{2} \sigma + \sigma^3  \, , \label{batchelorRelative}
\end{equation}
where $D_0 := \mu_0 \kBolt T$ is the diffusion coefficient.
The collective diffusion coefficients read
\begin{equation}
 \frac{D_{zz}^{\mathrm{C}}}{2D_0} = 1 + \frac{3}{4} \sigma + \frac{\sigma^3}{2}  \, , \quad 
 \frac{D_{xx}^{\mathrm{C}}}{2D_0} = 1 + \frac{3}{2} \sigma - \sigma^3  \, , \label{batchelorColective}
\end{equation}

\subsection{Self-diffusion for finite-sized particles}

From Eqs.~\eqref{eq:correlationFct}-\eqref{pairDiff} we first obtain the scaled self-diffusion coefficient for the motion of a single particle perpendicular to the membrane, 
\begin{equation}\label{DiffusionCoeff_Perp_Shear_Bending}
\begin{split}
  \frac{D_{zz}^{\mathrm{S}}}{D_0} &= 1-\frac{3}{32}\frac{\tauS(3B+2\tauS)}{(B+\tauS)^2} \epsilon 
					      + \frac{\tauS}{16} \frac{ 3\tauS^2+8B\tauS+6B^2}{(B+\tauS)^3} \epsilon^3   \\
					      &- \frac{\tauS}{64} \frac{4\tauS^3+15B\tauS^2+20B^2\tauS+10B^3}{(B+\tauS)^4} \epsilon^5  \, \\
                                              &- \frac{\epsilon }{12} \int_0^{\infty} \left(3+3u-\epsilon^2 u^2 \right)^2 \left( 1 - \frac{1-e^{-{\tauB u^3}}}{\tauB u^3} \right) e^{-2u} \Intd u \, ,
\end{split}
\end{equation}
where $\tauS := t/\TS$ and $\tauB := t/\TB$ are dimensionless times for shearing and bending, respectively.

For motion parallel to the membrane the scaled self-diffusion coefficient reads
\begin{equation}\label{DiffusionCoeff_Para_Shear_Bending}
\begin{split}
  \frac{D_{xx}^{\mathrm{S}}}{D_0} &= 1 - \frac{3}{64} \bigg( \frac{(2\tauS+3B)(5\tauS+4B)}{(\tauS+B)^2} - \frac{4B}{\tauS} \ln\left(1+\frac{\tauS}{B} \right)  \\
                                  &- \frac{16}{\tauS} \ln\left( 1+ \frac{\tauS}{2} \right) \bigg) \epsilon
			  + \frac{\tauS}{32} \frac{\tauS^2+3B\tauS+3B^2}{(\tauS+B)^3} \epsilon^3  \\
			  &-\frac{\tauS}{128} \frac{4\tauS^3+15B\tauS^2+20B^2\tauS+10B^3}{(\tauS+B)^4} \epsilon^5  \, \\
			 &- \frac{\epsilon}{12} \int_0^{\infty} \left( 3-\epsilon^2 u \right)^2 \left( u^2 -   \frac{1-e^{-{\tauB u^3}}}{\tauB u} \right) e^{-2u} \Intd u \, .
\end{split}
\end{equation}
We mention that Eqs.~\eqref{DiffusionCoeff_Perp_Shear_Bending} and \eqref{DiffusionCoeff_Para_Shear_Bending} correspond to leading order in $\epsilon$ to the ones reported in our earlier work \cite{daddi16}. For long times, the perpendicular velocity auto-correlation function $\phiPerpShearSelf$ decays as $t^{-4}$  whereas the bending part $\phiPerpBendingSelf$ as $t^{-4/3}$. For parallel motion, both the shearing and bending parts in the velocity auto-correlation function have a long-time tail of $t^{-2}$.






\subsection{Pair-diffusion for finite-sized particles}

The pair-diffusion coefficients are readily obtained by plugging Eqs.~\eqref{muZZ_Elastic} through \eqref{muXZ_Elastic} into Eqs.~\eqref{eq:correlationFct}-\eqref{pairDiff}:
\begin{widetext}
\begin{subequations}
\begin{align}
 \frac{D_{zz}^{\mathrm{P}} }{D_0} &= \frac{3\sigma}{4} + \frac{\sigma^3}{2}  - \frac{\sigma}{12 \xi^{5/2}} \int_{0}^{\infty}  \left( u \Lambda^2  \Pi_{\mathrm{S}} 
			  + \frac{2\Gamma_{-}^2}{ u^3} \Pi_{\mathrm{B}} \right) \chi_0 e^{-2u} \Intd u \, , \label{D_zz_P} \\
 \frac{D_{xx}^{\mathrm{P}} }{D_0} &= \frac{3\sigma}{2}-\sigma^3 -  \sigma \int_{0}^{\infty} \left( \frac{1}{24\xi^{5/2}} \left( -\xi^{1/2}\chi_1+2u\chi_0 \right)
			  \left( \Gamma_{+}^2 \Pi_{\mathrm{S}} + 2\Lambda^2 \Pi_{\mathrm{B}} \right)
			  +\frac{3 \chi_1}{2} \Pi_{\mathrm{S}}' 
			  \right) \frac{e^{-2u}}{u^2} \Intd u \, , \label{D_xx_P} \\
 \frac{D_{yy}^{\mathrm{P}} }{D_0} &= \frac{3\sigma}{4} + \frac{\sigma^3}{2} - \sigma \int_{0}^{\infty} \left( \frac{\chi_1}{24\xi^2}   \left( \Gamma_{+}^2 	\Pi_{\mathrm{S}} + 2\Lambda^2 \Pi_{\mathrm{B}} \right)
			   +\frac{3 \Pi_{\mathrm{S}}'}{2\xi^{1/2}}  \left( -\xi^{1/2}\chi_1+2u\chi_0 \right)
			  \right) \frac{e^{-2u}}{u^2} \Intd u \, , \label{D_yy_P} \\
 \frac{D_{xz}^{\mathrm{P}} }{D_0} &= \frac{\sigma}{12\xi^{5/2}} \int_{0}^{\infty} \left( \Gamma_{+}\Pi_{\mathrm{S}} + \frac{2\Gamma_{-} \Pi_{\mathrm{B}}}{u^2} \right) \chi_1 \Lambda e^{-2u}  \Intd u \, , \label{D_xz_P}
\end{align}
\end{subequations}
where we define
\begin{equation}
  \Pi_{\mathrm{S}} := \frac{B e^{-\frac{2u\tauS}{B}}+2u\tauS-B}{\tauS}  \, , \quad 
 \Pi_{\mathrm{S}}' := \frac{e^{-\tauS u}+\tauS u - 1}{\tauS} \notag  \, , \quad
 \Pi_{\mathrm{B}} := \frac{e^{-\tauB u^3}+\tauB u^3 - 1}{\tauB} \, . 
\end{equation}
\end{widetext}

We observe that the $xx$, $yy$ and $zz$ cross-correlation functions have the same large time behavior as their corresponding auto-correlation functions. For the component $\phi_{xz}^{\mathrm{P}}$, the shearing and bending related parts have large-time tails of $t^{-4}$ and $t^{-2}$, respectively.


\begin{figure}
  \begin{center}
     \includegraphics[scale = 0.35]{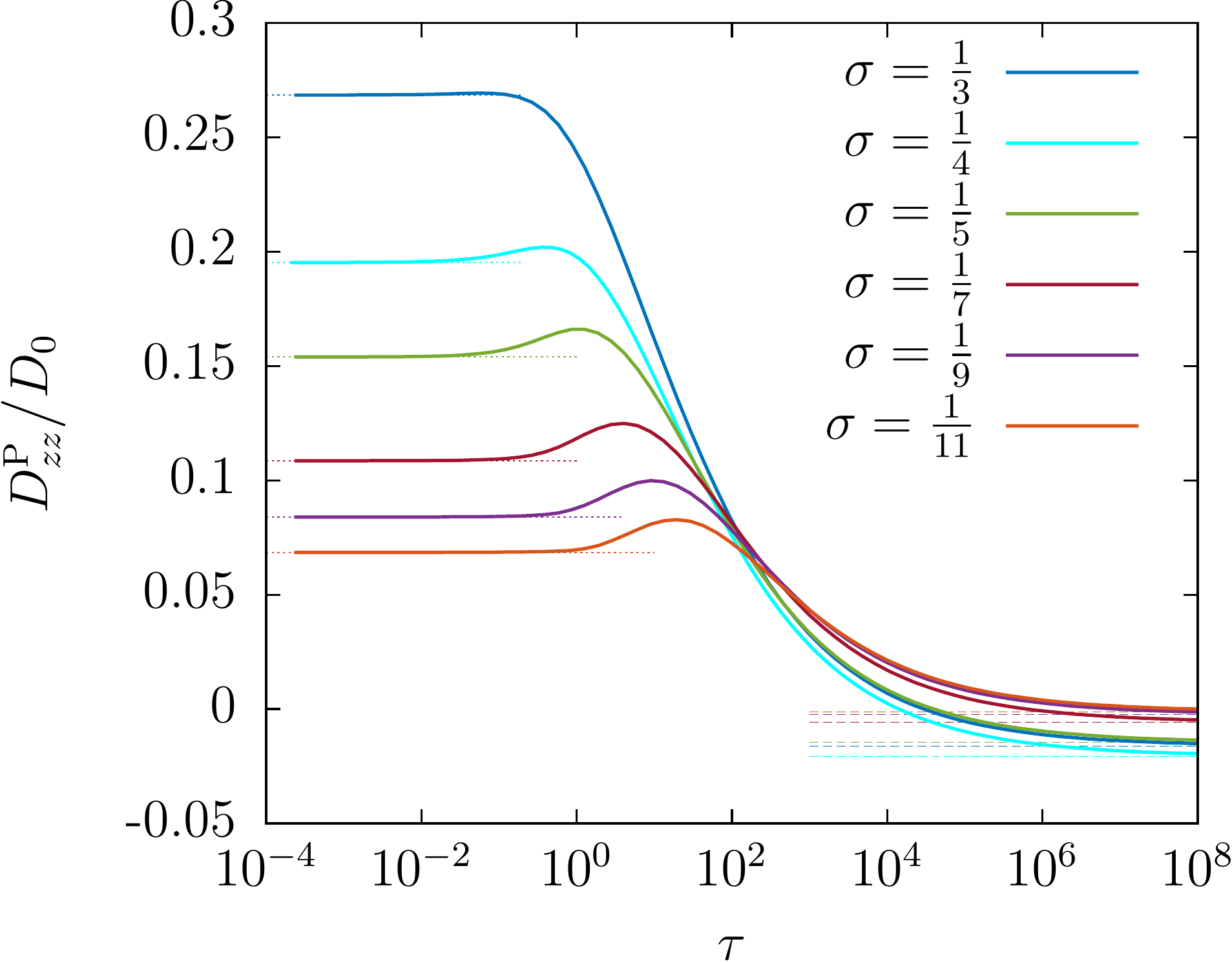}
     \caption{(Color online) The $zz$ component of the scaled pair-diffusion tensor versus the scaled time as given by Eq.~\eqref{D_zz_P} for different values of $\sigma$ with the parameters of Fig.~\ref{mobiCorr}.
     Horizontal dotted and dashed lines represent the bulk and hard-wall limits, respectively. For large inter-particle distances (small $\sigma$) a short superdiffusive regime is observed.
     } 
     \label{MSD_compare}
  \end{center}
\end{figure}

Fig.~\ref{MSD_compare} shows the variations of the $zz$ component of the scaled pair-diffusion coefficient as stated by Eq.~\eqref{D_zz_P} upon varying $\sigma$.
We observe that as $\sigma$ decreases, i.e. when the two particles stand further apart, the pair-diffusion coefficient can rise and exceed the bulk value for intermediate time scales as hinted on already by the pair-mobility around $\beta\sim1$ (cf. inset of Fig.~\ref{mobiCorr} $a$).
Such behavior is a clear signature of a short-lived superdiffusive regime.


\begin{figure}
  \begin{center}
     \includegraphics[scale = 0.4]{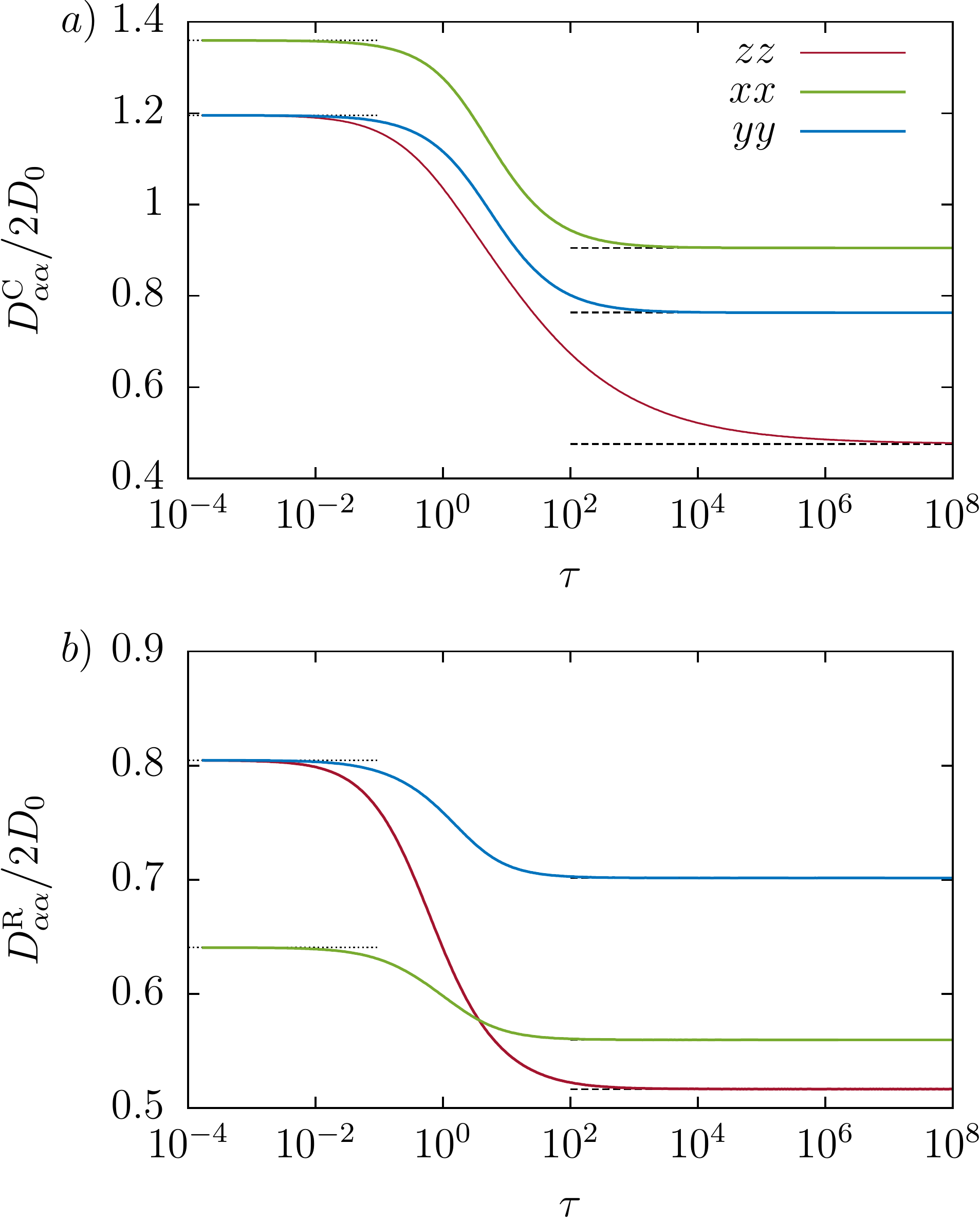}
     \caption{ (Color online) The scaled collective $(a)$ and relative $(b)$ diffusion coefficients as defined by Eq.~\eqref{collectiveRelativeModeDef} versus the scaled time.
     The horizontal dotted and dashed lines correspond to the bulk and hard-wall limits, respectively. 
     }
     \label{DiffCoefColectivefRelative}
  \end{center}
\end{figure}

In Fig.~\ref{DiffCoefColectivefRelative} we show the variations of the scaled collective and relative diffusion coefficients as defined by Eq.~\eqref{collectiveRelativeModeDef}, versus the scaled time $\tau$, using the parameters of Fig.~\ref{mobiCorr}.
At shorter time scales, the particle pair exhibits a normal bulk diffusion, since the motion is hardly affected by the presence of the membrane.
As a result, the diffusion coefficients are the same as calculated by Batchelor and given by Eq.~\eqref{batchelorRelative}.
As the time increases, both diffusion coefficients' curves bend down substantially to asymptotically approach the diffusion coefficients near a hard-wall.



\section{Conclusions} \label{sec:conclusions}

We have investigated the hydrodynamic interaction of a finite-size particle pair nearby an elastic membrane endowed with shear and bending rigidity.
Using multipole expansions together with \Faxen's law, we have provided analytical expressions for the frequency-dependent self- and pair-mobilities.
We have demonstrated that shearing and bending contributions may give positive or negative contributions to particle pair-mobilities depending on the inter-particle distance and the pair location above the membrane.
Most prominently, we have found that two particles approaching a membrane with only shearing resistance (as is typically assumed for elastic capsules) may experience hydrodynamic attraction in contrast to the well-known case of a hard wall where the interaction is repulsive. 
This unexpected effect will facilitate chemical reactions near the surface and may possibly even lead to the formation of particle clusters near elastic membranes.
On the other hand, membranes with bending resistance (such as vesicles) induce repulsive interactions similar to the hard wall. 
All our theoretical mobilities are validated by detailed boundary integral simulations.

Using the frequency-dependent particle mobilities, we have computed self- and pair-diffusion coefficients. Most commonly, relative and collective pair-diffusion is subdiffusive on intermediate time scales similar to earlier observations on the diffusion of a single particle \cite{daddi16}. A notable exception is the $zz$-component of the pair-mobility tensor which for certain parameters and frequencies surpasses its corresponding bulk value.
This induces a short-lasting superdiffusive regime in the corresponding mean-square-displacement.

\acknowledgments
The authors gratefully acknowledge funding from the Volkswagen Foundation 
as well as computing time granted by the Leibniz-Rechenzentrum on SuperMUC.

\appendix

\section{Derivation of Green's functions}\label{appendix:derivationGreenFcts}

In this appendix, we briefly sketch the derivation of the Green's functions in the presence of an elastic membrane, as stated by Eqs.~\eqref{Gzz} through \eqref{Gxz} of the main text.
For the solution of the steady Stokes equations Eqs.~\eqref{eq:momentum} and \eqref{eq:incompressibility},
we use a two-dimensional Fourier transform technique.
The variables $x$ and $y$  are transformed into the wavevector components $q_x$ and $q_y$.
Here we use the convention with a negative exponent for the forward Fourier transforms.
The transformed equations read
 \begin{align}
 -q^2 \vt_{x} + \vt_{x,zz} + i q_x \pt + \Ft_x \delta (z-z_0)  &= 0 \, , \notag \\
 -q^2 \vt_{y} + \vt_{y,zz} + i q_y \pt + \Ft_y \delta (z-z_0)  &= 0 \, , \notag \\
 -q^2 \vt_{z} + \vt_{z,zz} - \pt_{,z}  + \Ft_z \delta (z-z_0)  &= 0 \, , \notag \\
 -i q_x \vt_{x} - i q_y \vt_{y} + \vt_{z,z} &= 0 \, , \notag
 \end{align}
where a comma in indices denotes the spatial derivative with respect to the following coordinate.

We introduce a new orthonormal system in which the Fourier transformed vectorial quantities are decomposed into longitudinal, transverse and normal components, denoted by $\vt_l$, $\vt_t$ and $\vt_z$ respectively.
The corresponding orthonormal in-plane unit vector basis are
\begin{equation}
 \vect{q}_l := \frac{q_x}{q} \vect{e}_x + \frac{q_y}{q} \vect{e}_y \, , \quad
 \vect{q}_t := \frac{q_y}{q} \vect{e}_x - \frac{q_x}{q} \vect{e}_y \, , \label{transformation}
\end{equation}
where $q:=\sqrt{q_x^2 + q_y^2}$ is the wavenumber.
After transformation, the momentum equations become \cite{bickel07}
\begin{subequations}
      \begin{align}
       q^2 \vt_t - \vt_{t,zz}  &=  \frac{\Ft_t}{\eta}  \delta (z-z_0) \, , \label{eq:transverseEquation}\\
       \vt_{z,zzzz}  - 2q^2 \vt_{z,zz} + q^4 \vt_z  &=  \frac{q^2 \Ft_z}{\eta} \delta(z-z_0) \notag \\
		    &\quad +  \frac{iq \Ft_l}{\eta} \delta'  (z - z_0) \, , \label{eq:normalEquation}
      \end{align}
\end{subequations}
where $\delta'$ is the derivative of the Dirac delta function.
The longitudinal component $\vt_l$ is readily determined from  $\vt_z$  via the incompressibility equation \eqref{eq:incompressibility} such that
\begin{equation}
 \vt_l  =  \frac{i \vt_{z,z}}{q} \, . \label{eq:longitudinalFromNormal}
\end{equation}

According to the Skalak \cite{skalak73} and Helfrich \cite{helfrich73} models, the linearized tangential and normal traction jumps across the membrane are related to the membrane displacement field $\vect{u}$ at $z=0$ by \cite{daddi16}
\begin{subequations}
      \begin{align}
      [\sigma_{z\alpha}] &= -\frac{\kS}{3} \left( \Delta_{\parallel} u_\alpha + (1+2C) e_{,\alpha} \right) \,, \quad \alpha \in \{ x,y \} \, , \label{tangentialCondition}\\
      ~[\sigma_{zz}] &= \kB \Delta_{\parallel}^2 u_z \, , \label{normalCondition}
      \end{align}
\end{subequations}
where the notation $[w] := w(0^{+}) - w(0^{-})$ designates the jump of the quantity $w$ across the membrane. 
Here $C := \kA/\kS$ is a dimensionless number representing the ratio of the area expansion modulus to shear modulus, and $\kB$ is the membrane bending modulus.
$\Delta_{\parallel} := \partial_{,xx} + \partial_{,yy}$ denotes the Laplace-Beltrami operator along the membrane and $e := u_{x,x}+u_{y,y}$ is the dilatation function, mathematically  defined as the trace of the in-plane strain tensor.

The membrane displacement $\vect{u}$ as appearing in Eqs.~\eqref{tangentialCondition} and \eqref{normalCondition} is related to the fluid velocity by the no-slip boundary condition at the undisplaced membrane which reads
\begin{equation}
 \vt_\alpha = i\omega \ut_\alpha |_{z=0} \, .
\end{equation}

After solving the transformed equations \eqref{eq:transverseEquation}, \eqref{eq:normalEquation} and \eqref{eq:longitudinalFromNormal} and properly applying the boundary conditions at the membrane, we find that the diagonal components of the Green's function for $z \ge 0$ read
\begin{subequations}
\begin{align}
 \Gt_{zz}  &= \frac{1}{4 \eta q} 
 \bigg(
 \left( 1+q|z - z_0| \right) e^{-q|z-z_0|}  \notag \\
 & + \left( \frac{i\alpha z z_0 q^3}{1-i\alpha q} + \frac{i\alphaB^3 q^3 (1+qz)(1+q z_0)}{1-i\alphaB^3 q^3} \right) e^{-q(z+z_0)}  
 \bigg) \, ,
 \notag
 \\
  \Gt_{ll}  &= \frac{1}{4 \eta q} 
 \bigg(
(1-q |z - z_0|) e^{-q|z-z_0|}  \notag \\
 &+ \left( \frac{i\alpha q (1-q z_0)(1-qz)}{1-i\alpha q} + \frac{i z z_0 \alphaB^3 q^5}{1-i \alphaB^3 q^3} \right) e^{-q(z+z_0)}  
 \bigg) \, , 
 \notag
 \\
 \Gt_{tt}  &= \frac{1}{2 \eta q} \left( e^{-q|z-z_0|} + \frac{i B\alpha q}{2-i B \alpha q} e^{-q(z+z_0)}  \right) \, ,
 \notag
\end{align}
\end{subequations}
and the off-diagonal component $\Gt_{lz}$ reads
\begin{align}
 \Gt_{lz}  &= \frac{i}{4 \eta q} 
 \bigg(
-q (z - z_0) e^{-q|z-z_0|} \notag \\
 &+ \left( \frac{i\alpha z_0 q^2 (1-qz)}{1-i\alpha q} - \frac{i\alphaB^3 z q^4 (1+q z_0)}{1-i \alphaB^3 q^3} \right) e^{-q(z+z_0)}  
 \bigg) \, , 
 \notag
\end{align}
where $\alpha := \kS/(3 B\eta\omega)$ is a characteristic length scale for shearing and area expansion with $B := 2/(1+C)$, and $\alphaB := (\kB /(4\eta\omega))^{1/3}$ is a characteristic length scale for bending.
Furthermore, $\Gt_{tz} = \Gt_{zt} = 0$ because of the decoupled nature of Eqs.~\eqref{eq:transverseEquation} and \eqref{eq:normalEquation}.
Employing the transformation equations \eqref{transformation} back to the usual Cartesian basis, we obtain
\begin{subequations}
\begin{align}
 \Gt_{xx} (\vect{q},z,\omega) &=  \Gt_{ll} (q,z,\omega)\cos^2\phi +   \Gt_{tt} (q,z,\omega) \sin^2\phi \, \notag ,\\
 \Gt_{yy} (\vect{q},z,\omega) &=  \Gt_{ll} (q,z,\omega) \sin^2\phi+  \Gt_{tt} (q,z,\omega)\cos^2\phi \, , \notag \\
 \Gt_{xz} (\vect{q},z,\omega) &=  \Gt_{lz} (q,z,\omega) \cos \phi \, , \notag
\end{align}
\end{subequations}
where $\phi := \arctan (q_y/q_x)$.

The components $\Gt_{yz}$ and $\Gt_{zy}$ are irrelevant for our discussion because the resulting mobilities vanish, thus they are omitted here.
In addition, the component $\Gt_{zx}$ leads to the same mobility as $\Gt_{xz}$ because of the symmetry of the mobility tensor.
Furthermore, we define
\begin{equation}
 \Gt_{\pm} (q,z,\omega) := \Gt_{tt}(q,z,\omega) \pm \Gt_{ll}(q,z,\omega) \, . \notag
\end{equation}
Eqs.~\eqref{Gzz}-\eqref{Gxz} of the main text follow immediately after performing the two dimensional inverse spatial Fourier transform of the Green's function \cite{bracewell99}.


\section{Vanishing frequency behavior}\label{appendix:vanishingFrequencyLimits}

In the following, analytical expressions of the shearing and bending related parts in the particle self- and pair-mobilities are provided in the vanishing frequency limit.

\subsection{Self mobilities}

By taking the vanishing frequencies limit in Eqs.~\eqref{deltaMuPerpShear} and \eqref{deltaMuPerpBending}, 
the shearing and bending related corrections for the perpendicular motion read
  \begin{align}
     \lim_{\beta \to 0} \frac{\Delta \mu_{zz, \mathrm{S}}^{\mathrm{S}}}{\mu_0} &= -\frac{3}{16} \epsilon + \frac{3}{16} \epsilon^3 -\frac{1}{16} \epsilon^5 \, , 
     \notag \\
     \lim_{\betaB \to 0} \frac{\Delta \mu_{zz, \mathrm{B}}^{\mathrm{S}}}{\mu_0} &= -\frac{15}{16} \epsilon + \frac{5}{16} \epsilon^3 -\frac{1}{16} \epsilon^5  \, ,
     \notag
  \end{align}
leading to the hard-wall limit Eq.~\eqref{hardWallPerp} after summing up both contributions term by term.
Similarly, for the parallel motion, by taking the vanishing frequency limit in Eqs.~\eqref{deltaMuParaShear} and \eqref{deltaMuParaBending} we get
 \begin{align}
  \lim_{\beta \to 0} \frac{\Delta \mu_{xx,\mathrm{S}}^{\mathrm{S}}}{\mu_0} &= -\frac{15}{32} \epsilon + \frac{1}{32} \epsilon^3 -\frac{1}{32} \epsilon^5  \, ,  
  \notag \\
  \lim_{\betaB \to 0} \frac{\Delta \mu_{xx,\mathrm{B}}^{\mathrm{S}}}{\mu_0} &= -\frac{3}{32} \epsilon + \frac{3}{32} \epsilon^3 -\frac{1}{32} \epsilon^5  \, ,  
  \notag
 \end{align}
which also give the hard-wall limit Eq.~\eqref{hardWallPara} when summing up both parts.


\subsection{Pair mobilities}

By considering independently the shearing and bending related parts in the pair-mobility corrections as given by Eqs.~\eqref{muZZ_Elastic} through \eqref{muXZ_Elastic}, and taking the vanishing frequency limit, we obtain for the shearing part
\begin{subequations}
 \begin{align}
  \lim_{\beta\to 0} \frac{\Delta \mu_{zz, \mathrm{S}}^{\mathrm{P}}}{\mu_0} &= -\frac{3}{16}\frac{\xi(2\xi-1)}{(1+\xi)^{5/2}} \sigma
								  + \frac{3}{4}\frac{\xi(2\xi-3)}{(1+\xi)^{7/2}} \sigma^3 \notag \\
								  &-  \frac{2\xi^2-6\xi + \frac{3}{4}}{(1+\xi)^{9/2}} \sigma^5 \, , \label{muZZ_S} \\
  \lim_{\beta\to 0} \frac{\Delta \mu_{xx, \mathrm{S}}^{\mathrm{P}}}{\mu_0} &= -\frac{3}{16}\frac{5\xi^2+10\xi+8}{(1+\xi)^{5/2}} \sigma
								  + \frac{1}{4}\frac{\xi^2-10\xi+4}{(1+\xi)^{7/2}} \sigma^3 \notag \\
								  &- \frac{\xi^2- \frac{27}{4} \xi+1}{(1+\xi)^{9/2}} \sigma^5 \, , \label{muXX_S} \\
  \lim_{\beta\to 0} \frac{\Delta \mu_{yy, \mathrm{S}}^{\mathrm{P}}}{\mu_0} &= -\frac{3}{16}\frac{5\xi+4}{(1+\xi)^{3/2}} \sigma
								  + \frac{1}{4}\frac{\xi-2}{(1+\xi)^{5/2}} \sigma^3 \notag \\
								  &- \frac{\xi-\frac{1}{4}}{(1+\xi)^{7/2}} \sigma^5 \, , \label{muYY_S} \\
  \lim_{\beta\to 0} \frac{\Delta \mu_{xz, \mathrm{S}}^{\mathrm{P}}}{\mu_0} &= \frac{3}{16}\frac{(\xi-2)\xi^{1/2}}{(1+\xi)^{5/2}} \sigma 
								   -\frac{3}{4} \frac{(3\xi-2)\xi^{1/2}}{(1+\xi)^{7/2}} \sigma^3 \notag \\
								   &+\frac{5}{4} \frac{(4\xi-3)\xi^{1/2}}{(1+\xi)^{9/2}} \sigma^5 \, , \label{muXZ_S}
 \end{align}
\end{subequations}
and for the bending part
\begin{subequations}
 \begin{align}
  \lim_{\betaB\to 0} \frac{\Delta \mu_{zz, \mathrm{B}}^{\mathrm{P}}}{\mu_0} &= -\frac{3}{16}\frac{10\xi^2+11\xi+4}{(1+\xi)^{5/2}} \sigma
								  + \frac{1}{4}\frac{10\xi^2-7\xi-2}{(1+\xi)^{7/2}} \sigma^3 \notag \\
								  &-  \frac{2\xi^2-6\xi+\frac{3}{4}}{(1+\xi)^{9/2}} \sigma^5 \, , \label{muZZ_B} \\
  \lim_{\betaB\to 0} \frac{\Delta \mu_{xx, \mathrm{B}}^{\mathrm{P}}}{\mu_0} &= -\frac{3}{16} \frac{\xi(\xi-2)}{(1+\xi)^{5/2}} \sigma
								   +\frac{3}{4} \frac{\xi(\xi-4)}{(1+\xi)^{7/2}} \sigma^3 \notag \\
								   &-\frac{\xi^2-\frac{27}{4}\xi+1}{(1+\xi)^{9/2}} \sigma^5 \, , \label{muXX_B} \\
  \lim_{\betaB\to 0} \frac{\Delta \mu_{yy, \mathrm{B}}^{\mathrm{P}}}{\mu_0} &= -\frac{3}{16} \frac{\xi}{(1+\xi)^{3/2}} \sigma
								    +\frac{3}{4} \frac{\xi}{(1+\xi)^{5/2}} \sigma^3 \notag \\
								    &- \frac{\xi-\frac{1}{4}}{(1+\xi)^{7/2}} \sigma^5 \, , \label{muYY_B} \\
  \lim_{\betaB\to 0} \frac{\Delta \mu_{xz, \mathrm{B}}^{\mathrm{P}}}{\mu_0} &=	\frac{3}{16} \frac{(5\xi+2)\xi^{1/2}}{(1+\xi)^{5/2}} \sigma
								    -\frac{15}{4} \frac{\xi^{3/2}}{(1+\xi)^{7/2}} \sigma^3 \notag \\
								    &+\frac{5}{4} \frac{(4\xi-3) \xi^{1/2}}{(1+\xi)^{9/2}} \sigma^5 \, . \label{muXZ_B}
 \end{align}
\end{subequations}

The total correction as given by Eqs.~\eqref{muZZ} through \eqref{muXZ} is recovered by summing up term by term both contributions.


%

\end{document}